\pdfoutput=1

\documentclass[aps,prb,reprint,showpacs,superscriptaddress]{revtex4-1}

\usepackage{graphicx}
\usepackage{graphics}
\usepackage{amsmath}
\usepackage{amssymb}
\usepackage{amsfonts}
\usepackage{dcolumn}
\usepackage{dsfont}
\usepackage{latexsym}
\usepackage{rotating}
\usepackage{color}
\usepackage{latexsym}
\usepackage{bbm}
\usepackage{subfigure}
\usepackage{float}
\usepackage{epsfig}
\usepackage{epsf}
\usepackage{psfrag}
\usepackage{bm}
\usepackage{amsthm}
\usepackage{eucal}
\usepackage{mathrsfs}
\usepackage{url}
\usepackage{braket}

\usepackage{color} % use colored text in latex

\newcommand{\mn}{MnV$_2$O$_4$}
\newcommand{\md}{\mathrm{d}}
\usepackage{hyperref}
\hypersetup{
colorlinks=true,final=true,
        linkcolor=blue,
        citecolor=blue,
        filecolor=blue,
        urlcolor=blue,
}
\begin{document}
\title{Phonon dispersion, Raman spectra and evidence for spin-phonon coupling in MnV$_2$O$_4$ from first-principles}

\author{Dibyendu Dey}
\email{ddey3@asu.edu}
\affiliation{Department of Physics, Indian Institute of Technology Kharagpur, Kharagpur - 721302, India}
\affiliation{Department of Physics, Arizona State University, Tempe, AZ - 85287, USA}

\author{T. Maitra}
\affiliation{Department of Physics, Indian Institute of Technology Roorkee, Roorkee - 247667, India}

\author{U. V. Waghmare}
\affiliation{Theoretical Sciences Unit, Jawaharlal Nehru Centre for Advanced Scientific Research, P.O. Jakkur, Bangalore 560064, India}

\author{A. Taraphder}
\affiliation{Department of Physics, Indian Institute of Technology Kharagpur, Kharagpur - 721302, India}
\affiliation{Centre for Theoretical Studies, Indian Institute of Technology Kharagpur, Kharagpur - 721302, India}
\affiliation{School of Basic Sciences, Indian Institute of Technology Mandi, Himachal Pradesh 175001, India}
\date{\today}

\begin{abstract}
MnV$_2$O$_4$ in the spinel structure is known to exhibit coupled orbital and spin ordering, and its Raman spectra show interesting anomalies in its low-temperature phase. With a goal to explain this behavior involving coupled spins and phonons, we determine here the spin-phonon couplings in MnV$_2$O$_4$ from a theoretical analysis of its phonon spectra and their dependence on spin-ordering and electron correlations, obtained from first-principles density functional theoretical calculations. Using these in an analysis based on a Landau-like theory, we uncover the mechanism governing the Raman anomalies observed in its low-temperature phase. 
\end{abstract}

\maketitle

\section{Introduction} 
Correlated electronic systems with spinel structure have seen a surge in interest of late, due inherently to rich physics involving frustrated magnetism, orbital ordering, charge ordering, etc. Fascinating physical phenomena emerge from the coupling between various degrees of freedom like spin, orbital and lattice, in the presence of both geometric frustration and Coulomb correlation~\cite{TokuraSc, LeeJPS, RadaNJP, KhomPRL, KatoPRL, TsunePRB}. The competition among different degrees of freedom manifests itself in several structural and magnetic transitions accompanied by orbital ordering in these systems~\cite{DebPRB, LeePRL, MacPRB, GarleaPRL}. These compounds, having generic formula AB$_2$O$_4$, consist of two distinct cation sites: A site is tetrahedrally coordinated with neighboring O sites while B site is octahedrally coordinated with O ligands~\cite{NiiPRB}, as shown schematically in Fig.~\ref{Fig1}(a). Geometrically frustrated magnetic spinels provide a fertile playground for exploration of interplay between these degrees of freedom~\cite{TchernyPRL, MatteoPRB}.  

MnV$_2$O$_4$ is a member of AB$_2$O$_4$ family, which has the A-site (Mn$^{+2}$, S=5/2) magnetic but not orbitally active, while the B-site (V$^{+3}$, S=1) is both magnetic and orbitally active. The combined effect of magnetic and orbital degrees brings in rich and complex physics. This compound undergoes a magnetic transition from paramagnetic (PM) to collinear ferrimagnetic (FiM) phase at T$_{N1}$= 57K; followed by a structural transition (cubic$\rightarrow$ tetragonal) together with a second magnetic transition at T$_{N2}$= 54K where non collinear FiM spin ordering sets in~\cite{GarleaPRL,SuzukiPRL,AdachiPRL}.

Earlier, the controversies mostly centered around the ordering of orbitals of V, and stimulated an upsurge of research in this system~\cite{GarleaPRL, AdachiPRL, SarkarPRL}. Various possible scenarios were proposed for the orbital order based on the symmetry of the tetragonal phase and the type of dominant interactions~\cite{GarleaPRL, AdachiPRL}. In a recent theoretical study~\cite{DeyPRB}, A-type antiferro-orbital ordering has been observed at the V sites where one t$_{2g}$ electron occupies the d$_{xy}$ orbital at every V site and other electron occupies d$_{xz}$ and d$_{yz}$ orbitals alternately along c direction. These findings are in good agreement with the experimental measurements in MnV$_2$O$_4$~\cite{GarleaPRL}.

Though much attention has been given to the electronic properties, the phonon-related phenomena and their coupling with other degrees of freedom remained unexplored. However, few fascinating experiments performed in the recent past indicate that phonons play an important role at the microscopic level. Takubo {\it et al.}~\cite{TakuboPRB}, investigated Raman scattering of MnV$_2$O$_4$ and observed that several peaks evolve in the Raman spectra below T$_{N2}$. They showed a peculiar polarization-dependence of B$_g$ modes, and tried to explain it with Mott excitation phenomena. However, the appearance of experimental intensity peak with XY polarization was not supported by their analysis. Their mode-assignment of the Raman-active peaks in the tetragonal phase was based on the I4$_1$/amd space group, whereas the space group symmetry of MnV$_2$O$_4$ is I4$_1$/a~\cite{GarleaPRL}. Later, an inelastic light scattering study of the temperature and magnetic field dependences of one-and two-magnon excitations in MnV$_2$O$_4$ revealed that spin-lattice coupling is indeed significant in its low-temperature phase~\cite{GleasonPRB}. Additionally, anomalous temperature-dependence of the peak intensity was observed which was attributed to a strong coupling between magnetic and vibrational excitations in MnV$_2$O$_4$. 

In view of the significant effects of phonons in this system, we have investigated the spin-phonon coupling across the Brillouin zone (BZ), along with Raman scattering to understand the polarization-dependence of B$_g$ modes from density functional theory (DFT) calculations. The paper is organized as follows. In Sec.~\ref{meth}, we briefly discuss the methodology used to calculate phonons and related properties. In Sec.~\ref{res}, first, we discuss the structural properties in detail and calculate phonon modes at $\Gamma$ point of the BZ. Next we present analysis Raman scattering intensities of the B$_g$ mode. Later, we study the effects of spin-phonon coupling and calculate the coupling term at different high symmetry points of the BZ and compare our results with experimental data. Finally, we give a brief summary and outlook in Sec.~\ref{conc}.

\section{Methods}
\label{meth}
Our first-principles calculations based on the density functional theory~\cite{KohnShamPR, HohenKohnPR} have been performed using plane-wave basis, as implemented in the Vienna {\it ab-initio} simulation package VASP~\cite{KresseVASP1, KresseVASP2}. We have used projector-augmented wave (PAW)~\cite{BlochPAW, KressePAW} potentials in our calculations and the wave functions were expanded in the plane-wave basis with a kinetic energy cutoff of 500 eV. We used a generalized gradient approximation (GGA) with the Perdew-Burke-Ernzerhof (PBE)~\cite{GGAPBE} parametrization for the exchange-correlation energy functional. Total energies were converged to less than 10$^{-8}$ eV to achieve self-consistency and Brillouin zone integration was sampled on a $\Gamma$ centered k-mesh of 6 $\times$ 6 $\times$ 6. Electron correlation effects beyond GGA were incorporated for 3$d$-electrons of Mn, and V ions within GGA+U~\cite{DudarevPRB} approximation where U is the on-site Coulomb correlation~\cite{Hubbard} (U = 5 eV for both the transition metals), and Hund’s coupling strengths (J) were set at J = 1 eV. In structural relaxation, positions of the ions were relaxed towards the equilibrium using conjugate gradient algorithm, until the Hellman-Feynman forces became less than 10$^{-3}$ eV/\AA. 

Dynamical matrix and phonons were calculated from a frozen-phonon method with atomic displacements of $\pm$ 0.03\AA, as driven by the PHONOPY code~\cite{Phonopy} and Raman scattering intensity peaks were obtained from PHONON software~\cite{PHONON} interfacing with DFT as implemented in the VASP code. Spin-phonon couplings were determined based on the scheme~\cite{RayPRB} used by Ray {\it et al.}.

\section{Results and Discussion}
\label{res}
\subsection{Structural properties}
\begin{figure}[ht]
%\vspace{-1.0cm}
\begin{center}
\includegraphics[width=9.0cm]{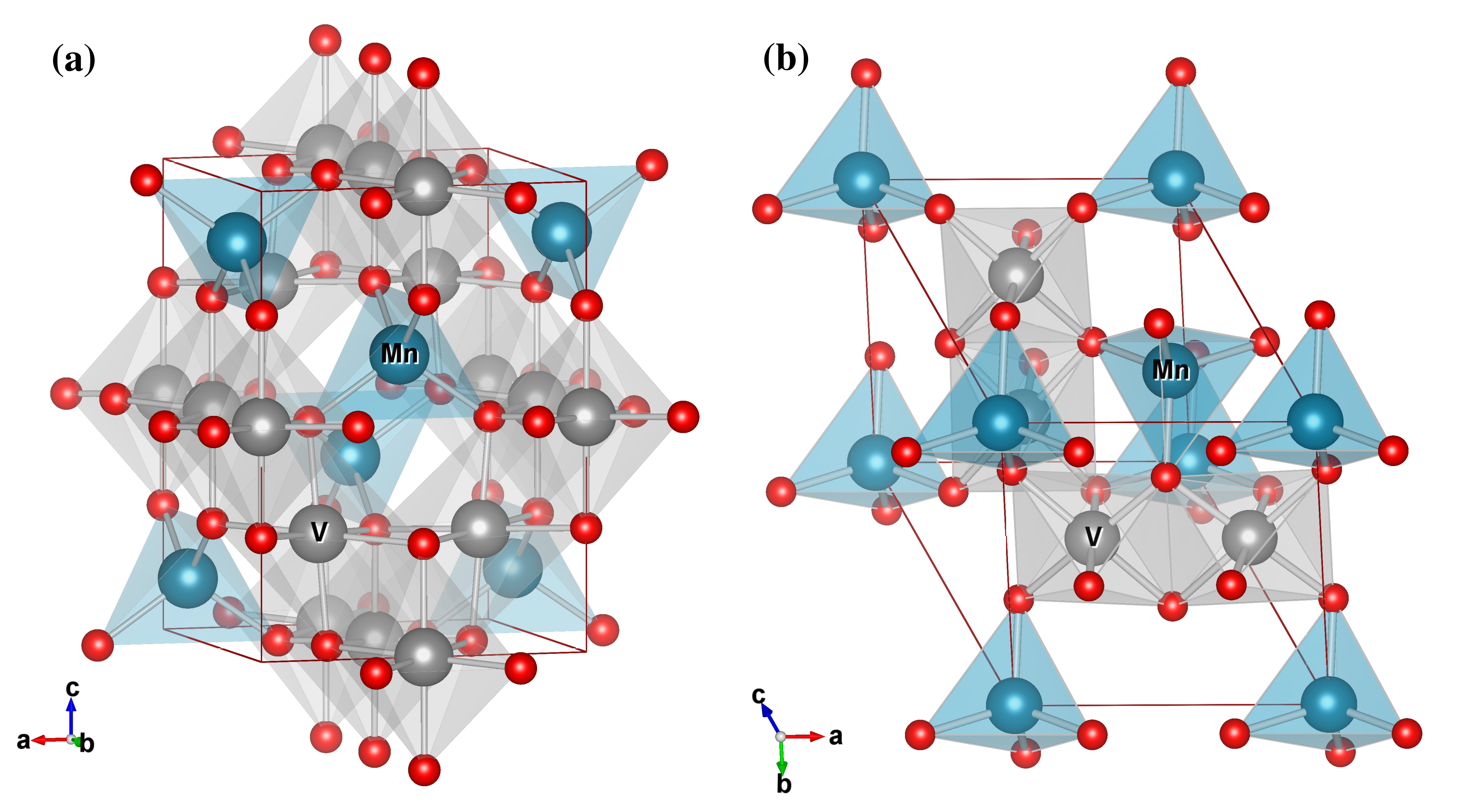}
\caption{(Color online) (a) Crystallographic unit cell and (b) primitive cell of MnV$_2$O$_4$ compound. Oxygen octahedra (tetrahedra) around V (Mn) ions are shown in grey (cyan-blue).}
\label{Fig1}
\end{center}
\end{figure}

The MnV$_2$O$_4$ compound, in its low-temperature phase, has a tetragonal structure with I4$_1$/a space group symmetry. A network of edge-shared VO$_6$ octahedra with interstitial MnO$_4$ tetrahedra form a geometrically frustrated lattice in which the mirror or glide plane perpendicular to the ab plane is absent. The tetragonal unit cell of MnV$_2$O$_4$ is shown in Fig.~\ref{Fig1}(a). 
%%%%%%%%%%%%%%%%%%%%%%%%%%%%%%%%%%%%%%%%  
%%%%%%%%%%%%%%%%%%%%%%%%%%%%%%%%%%%%%%%

\begin{figure}
%\vspace{-1.0cm}
\begin{center}
\includegraphics[width=7.2cm]{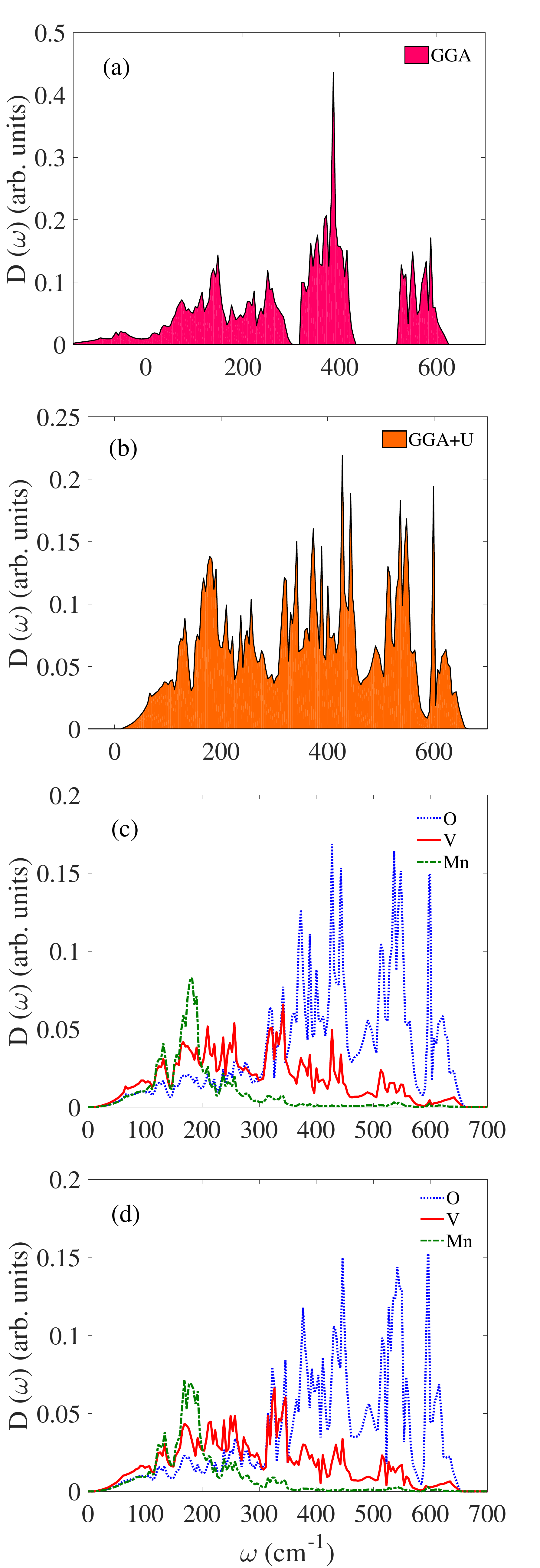}
\caption{\label{Fig2}(Color online) Total phonon-DOS is shown for (a) GGA, and (b) GGA+U calculations in FiM configuration. Unstable modes with imaginary frequencies (given as -ve) are observed only in GGA calculations. Projected DOS of Mn, V and O atoms within GGA+U calculations in (c) FiM, and (d) FM states.}
\end{center}
\end{figure}

\begin{figure}
%\vspace{-1.0cm}
\begin{center}
\includegraphics[width=9.2cm]{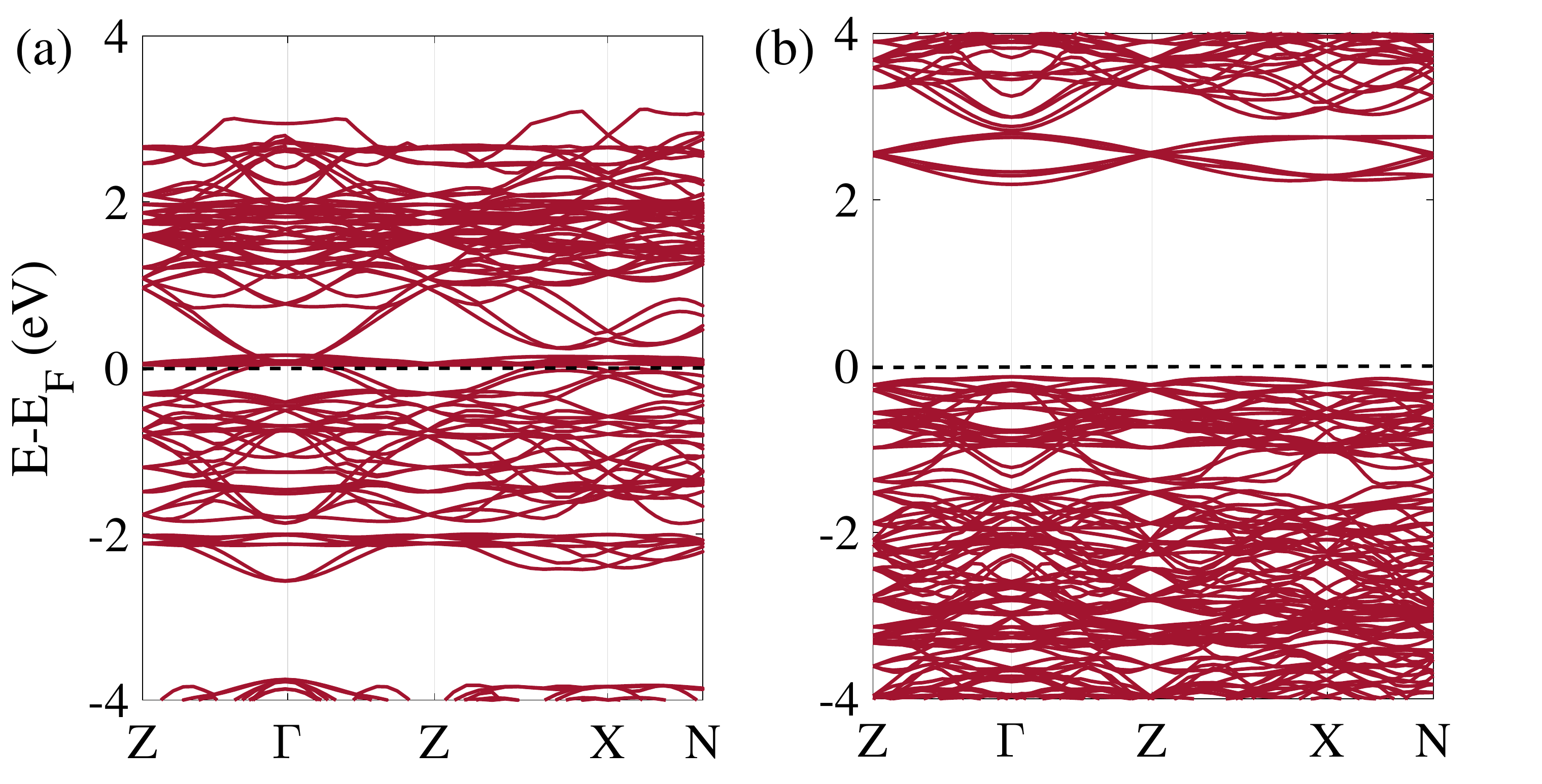}
\caption{\label{Fig3}(Color online) Electronic band structure of MnV$_2$O$_4$ in FiM state within (a) GGA and (b) GGA+U calculations.}
\end{center}
\end{figure}

\begin{figure}
%\vspace{-1.0cm}
\begin{center}
\includegraphics[width=7.5cm]{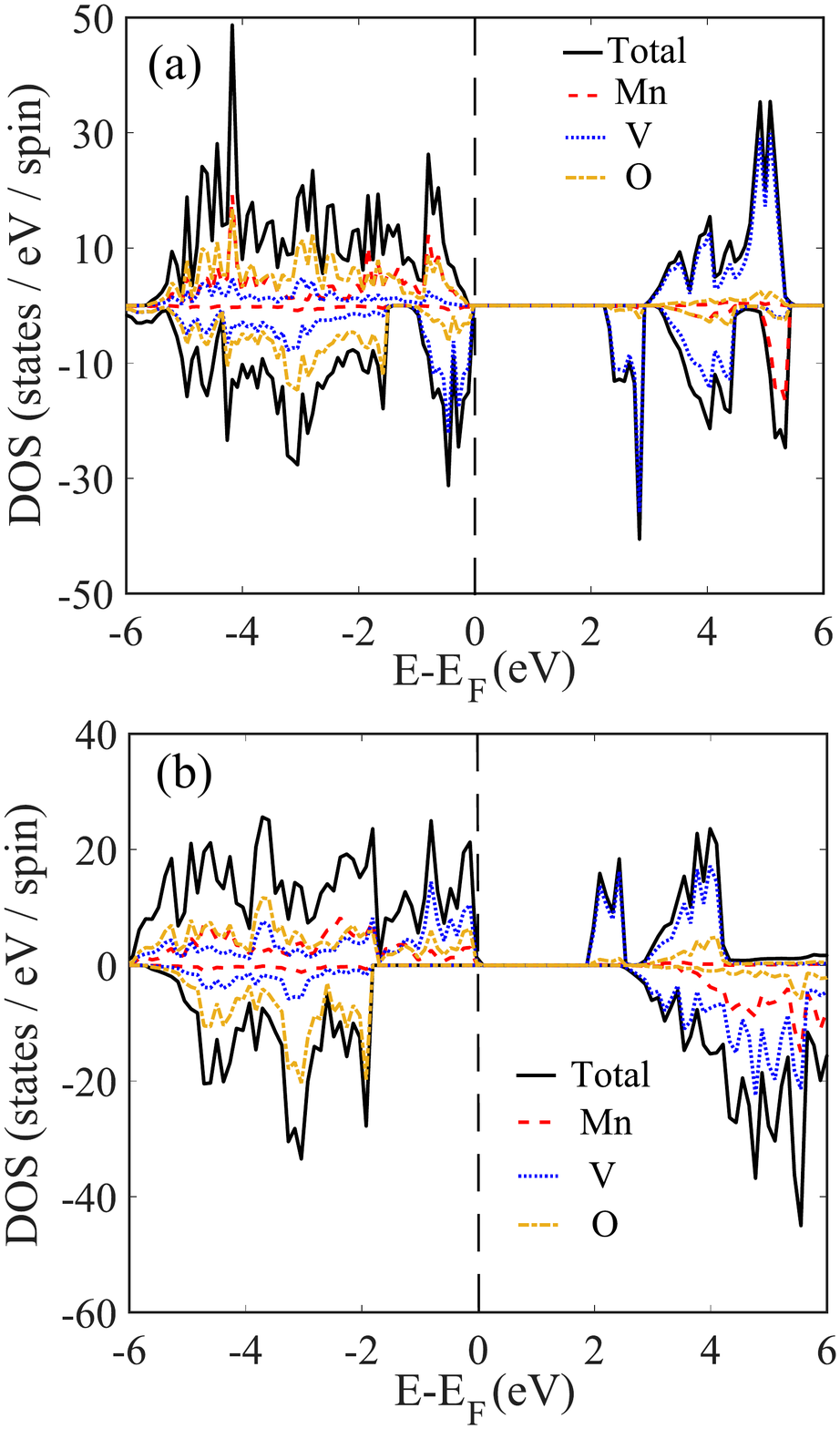}
\caption{\label{Fig4}(Color online) Total and projected electronic DOS of Mn, V and O atoms of MnV$_2$O$_4$ in (a) FiM, and (b) FM states within GGA+U calculations. DOS in the positive Y-axis represent up-spin states, while DOS in the negative Y-axis represent down-spin states}
\end{center}
\end{figure}

The lattice constants and the atomic positions are optimized within GGA and GGA+U approximations to achieve the minimum energy structure. However, the phonon calculations with the GGA-optimized structure show modes with imaginary frequencies in the phonon density of states (DOS) (Fig.~\ref{Fig2}(a)), which may arise due to the inaccuracy in calculating forces within GGA approximation. As the compound MnV$_2$O$_4$ is known to be a Mott insulator, strong correlation physics is incorporated in the calculations within GGA+U approximation. The calculations have been performed for U$_{eff}$ ($=U-J$, where U is the on-site Coulomb interaction and J is the Hund's exchange interaction) ranging from 3 eV to 5 eV at the Mn and V sites. For U$_{eff}$ $\geqslant$ 4 eV we observe no imaginary modes in the DOS (Fig.~\ref{Fig2}(b)). We would like to note that, our calculations do not change significantly for U$_{eff}$ above that value. Therefore, we present the results obtained for U$_{eff}=$4 eV at both sites. In addition, we have considered collinear ferrimagnetic (FiM) ordering of spins in our calculations where V spin moments are aligned opposite to Mn spin moments. Within the FiM spin configuration, the crystal structure preserves the symmetry of the experimental structure. It is evident from Fig.~\ref{Fig3}(a) that, the electronic band structure of MnV$_2$O$_4$ shows zero band gap within GGA calculations. On the other hand, GGA+U calculated electronic band structure (Fig.~\ref{Fig3}(b)) correctly represents the insulating state. We have also shown the atom projected electronic DOS (Fig.~\ref{Fig4}(a), (b)) for FiM and ferromagnetic (FM) spin states within GGA+U approximation.  In both cases, the DOS shows the insulating state.
%%%%%%%%%%%%%%%%%%%%%%%%%%%%%%%%%%%
%%%%%%%%%%%%%%%%%%%%%%%%%%%%%%%%%%%

%%%%%%%%%%%%%%%%%%%%%%%%%%%%%%%%%%%
\begin{table*}
\centering
\begin{tabular}{p{2.0cm}|p{3.5cm}|p{3.5cm}|p{3.5cm}|p{3.5cm}}
 \hline
  & \centering Experimental & \centering GGA & \centering GGA+U & ~~~~~~~~~~ GGA+U \\
  & & \centering (FiM) & \centering (FiM) & ~~~~~~~~~~~~ (FM)\\
 \hline
 \hline
Lattice Const. & \centering a=6.05\AA & \centering a=6.00\AA & \centering a=6.19\AA &~~~~~~~~~~ a=6.20\AA \\
  & \centering c=8.46\AA & \centering c=8.48\AA & \centering c=8.60\AA &~~~~~~~~~~ c=8.61\AA \\
%  & \centering c/a=0.99 & c/a=0.999 & \centering c/a=0.98 & \\
 \hline 
\centering Mn & \centering 0.000~~ 0.750~~ 0.875 & \centering 0.000~~ 0.750~~ 0.875  & \centering 0.000~~ 0.750~~ 0.875  & ~~~ 0.000~~ 0.750~~ 0.875 \\

\centering V & \centering 0.250~~ 0.250~~ 0.750 & \centering 0.250~~ 0.250~~ 0.750 &  \centering 0.250~~ 0.250~~ 0.750 & ~~~ 0.250~~ 0.250~~ 0.750 \\

\centering O & \centering 0.997~~ 0.474~~ 0.737 & \centering 0.999~~ 0.478~~ 0.739 & \centering 0.993~~ 0.474~~ 0.739 & ~~~ 0.993~~ 0.474~~ 0.739\\ 
 \hline
 \hline
\end{tabular}
\caption{Energy-minimized structural parameters of MnV$_2$O$_4$ in the FiM and FM states within GGA and GGA+U approximation. Experimental structural parameters are given for comparison\cite{NiiPRB}.}
\label{Tab1}
\end{table*}
%%%%%%%%%%%%%%%%%%%%%%%%%%%%%%%%%%%

In Table~\ref{Tab1} we compare the structural parameters of the GGA- and GGA+U-optimized structures for different spin configurations with the experimental one. We can see that the lattice constants, c/a ratio and the atomic positions of the optimized structures are comparable.
%%%%%%%%%%%%%%%%%%%%%%%%%%%%%%%%%%%
%%%%%%%%%%%%%%%%%%%%%%%%%%%%%%%%%%%
%%%%%%%%%%%%%%%%%%%%%%%%%%%%%%%%%%%
The MnV$_2$O$_4$ compound in the tetragonal phase has C$_{4h}$ point group symmetry and its primitive unit cell (See Fig.~\ref{Fig1}(b)) contains six magnetic ions (two Mn and four V) and eight non-magnetic O ions. Phonons at the $\Gamma$ point can be classified according to the irreducible representations of the C$_{4h}$ point group of \mn as $\Gamma$ = 6E$_u$ + 6A$_u$ + 6B$_u$ + 4B$_g$ + 4E$_g$ + 3A$_g$, where the acoustic modes are omitted. Among them, the B$_g$, E$_g$, and A$_g$ modes are Raman active, while the E$_u$ and A$_u$ modes are IR active.

\subsection{Raman activity} In this section, we calculate the Raman spectra from density functional theory. The irreducible representation of Raman-active modes at $\Gamma$ point are listed in Table~\ref{Tab2} along with the calculated phonon frequencies. Calculated Raman allowed phonon modes are in good agreement with the experiments~\cite{TakuboPRB}.
%%%%%%%%%%%%%%%%%%%%%%%%%%%%%%%
\begin{table}[ht]
\begin{center}
%\captionof{table}{}\label{TT1}
\begin{tabular}{p{1.5cm} |p{1.5cm} |p{1.5cm} |p{1.5cm} |p{1.5cm}} 
 \hline
\centering Modes & \centering Expt. (T=5K) & \centering FiM (GGA+U) & \centering FM (GGA+U) & ~~$\Delta_{rel}$(\%)\\ 
  \hline
  \hline
\centering  4B$_g$ & \centering $-$ & \centering 181 & \centering 184 & ~~ -1.7 \\
\centering   & \centering $-$ & \centering 367 & \centering 374 & ~~ -1.9 \\
\centering   & \centering 479 & \centering 440 & \centering 438 & ~~~ 0.4\\
\centering   & \centering 585 & \centering 539 & \centering 538 & ~~~ 0.2\\  
  \hline
\centering  4E$_g$ & \centering $-$ & \centering 185 & \centering 187 & ~~ -1.1 \\ 
\centering  & \centering $-$ & \centering 257 & \centering 248 & ~~~ 3.5 \\
\centering  & \centering $-$ & \centering 451 & \centering 449 & ~~~ 0.4 \\
\centering  & \centering 570 & \centering 549 & \centering 550 & ~~ -0.2 \\ 
  \hline 
\centering  3A$_g$ & \centering $-$ & \centering 319 & \centering 322 & ~~ -0.9 \\
\centering  & \centering $-$ & \centering 378 & \centering 377 & ~~~ 0.3 \\
\centering  & \centering 673 & \centering 633 & \centering 623 & ~~~ 1.6 \\  
  \hline
  \hline   
\end{tabular}
\caption{Irreducible representations of the Raman-active modes, calculated phonon frequencies (in cm$^{-1}$) for FiM and FM spin configurations in comparison with the Raman-active modes observed experimentally. The dashes in the second column imply that the intensity of the corresponding peak was too weak to observe experimentally~\cite{TakuboPRB}. Relative angular frequency shift ($\Delta_{rel}$=$\frac{\omega_{FiM}-\omega_{FM}}{\omega_{FiM}}\times$100\%) due to change in magnetic ordering of the Raman-active modes are given in the fifth column.}
\label{Tab2}
\end{center}
\end{table}
%%%%%%%%%%%%%%%%%%%%%%%%%%%%%%%%%%%%%%%%%%%%%%%%%%%%%%%%%%%

In a Raman scattering process, an incident photon of frequency $\omega_{l}$ and polarization versor {\textbf{g}}$_{l}$ either creates (Stokes process) or annihilates (anti-Stokes process) a phonon of frequency $\omega_{j}$ and scatters to an outgoing photon of frequency $\omega_{s}$ and polarization versor {\textbf{g}}$_{s}$. 
From the principle of energy conservation, we can write $\omega_{s}$ = $\omega_{l}$ $\pm$ $\omega_{j}$, where the (plus) minus sign ascribe to the (anti-) Stokes process. The differential cross section for Raman scattering in non-resonant conditions of the Stokes process involving a phonon of eigenmode $j$ is given by the following equation (for a unit volume of the sample)~\cite{Cardona, Bru}:
\begin{equation}
\frac{\md^2 \sigma}{\md \Omega \md \omega}=\sum\limits_{j}\frac{\omega_{s}^{4}}{c^4}\lvert \mathrm{\textbf{g}}_{s}.\overline{R}^{j}.\mathrm{\textbf{g}}_{l} \rvert^2[n_{b}(\omega)+1]\delta(\omega-\omega_{j})
\label{eq1}
\end{equation}
In this expression $n_{b}(\omega)$ is the Bose occupation factor, and c is the velocity of light in vacuum. The second rank tensor $\overline{R}^{j}$ in Eq.~\ref{eq1} is known as the Raman tensor associated with the phonon eigenmode $j$ which is given by the following equation: 
\begin{equation}
R_{\alpha\beta}^{j}=\sqrt{\frac{V\hslash}{2\omega_{j}}}\sum\limits_{\kappa=1}^{N}\frac{\partial \chi_{\alpha\beta}}{\partial \mathrm{\textbf{r}}(\kappa)}.\frac{\mathrm{\textbf{e}}(j,\kappa)}{\sqrt{M_{\kappa}}} ; (\alpha,\beta = 1, 2, 3)
\label{eq2}
\end{equation}
Here, $V$ is the unit cell volume. $\textbf{r}(\kappa)$ and $M_{\kappa}$ are the position and mass of the $\kappa^{th}$ atom respectively and the summation runs over all the $N$ atoms in the unit cell. The eigenstates and eigenvalues of the dynamical matrix at the $\Gamma$ point are denoted by $\textbf{e}(j,\kappa)$, and ${\omega_{j}}$ respectively. In Eq.~\ref{eq2}, the electric polarizability tensor $\chi_{\alpha\beta}$ is defined as: $\chi_{\alpha\beta}=\frac{1}{4\pi}(\epsilon_{\alpha\beta}-\delta_{\alpha\beta})$, where $\epsilon_{\alpha\beta}$ is the dielectric tensor. The tensor $\overline{R}^{j}$ is computed from the electric polarizability tensor within finite difference approach by moving the atoms of different symmetry with a displacement of 0.03\AA~\cite{UmariPRB,CeriottiPRB}.
%%%%%%%%%%%%%%%%%%%%%%%%%%%%%%%%%%%%%%%%%%%%%%%%%%%
\begin{figure}
%\vspace{-1.0cm}
\begin{center}
\includegraphics[width=9.0cm]{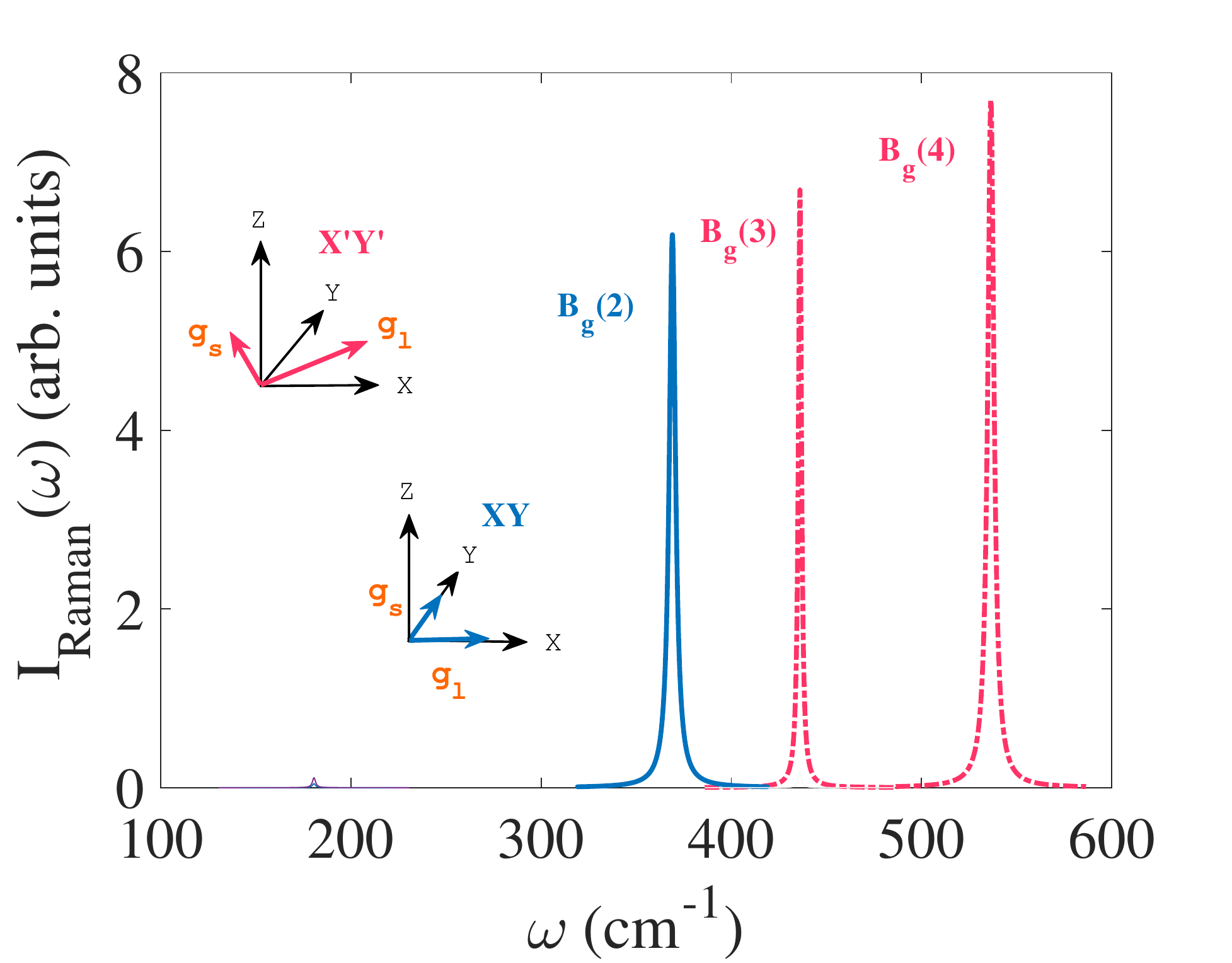}
\caption{\label{Fig5}(Color online) Calculated Raman spectrum for light
in the X$^{\prime}$Y$^{\prime}$ and XY polarization. Raman peaks with red (blue) color are observed in X$^{\prime}$Y$^{\prime}$ (XY) polarization.}
\end{center}
\end{figure}
%%%%%%%%%%%%%%%%%%%%%%%%%%%%%%%%%%%%%%%%%%%%%%%%%%%

In order to compare our theoretical results with the available experimental data, we have computed the intensity of the Raman spectra of a single-crystal sample from differential scattering cross-section as implemented in the PHONON software~\cite{PHONON} at various polarization configurations of the incident and scattered light (same as the experimental ones~\cite{TakuboPRB}). In our calculations, we have considered linear polarization along [100], [010], [110], and [1$\overline{1}$0] direction which is ascribed as X, Y, X$^{\prime}$, and Y$^{\prime}$ respectively. Temperature-dependence of the Raman spectra within the XY (polarization direction of the incident and scattered light respectively) and X$'$Y$'$ polarization configurations has been reported experimentally, and by analyzing the intensity peaks of certain modes, the nature of the orbital ordering in this compound was established\cite{TakuboPRB}. However, the phonon frequencies were calculated by assuming short-range force constants for the nearest- neighbor bonds and the modes were assigned (irreducible representation) based on the I4$_1$/amd space group, while the space group symmetry of the parent compound is actually I4$_1$/a. 

Calculated frequencies of the four Raman-allowed B$_g$ modes within I4$_1$/a space group symmetry are at 181cm$^{-1}$(B$_{g}$(1)), 367cm$^{-1}$(B$_{g}$(2)), 440cm$^{-1}$(B$_{g}$(3)), and 539cm$^{-1}$(B$_{g}$(4)). From our Raman scattering calculations, we found finite intensity peaks in the X$'$Y$'$ polarization (Fig.~\ref{Fig5}) at two higher frequency B$_{g}$ modes (B$_{g}$(3) and B$_{g}$(4)), while the lower modes (B$_{g}$(1) and B$_{g}$(2)) are absent. This result agrees well with the experimental findings~\cite{TakuboPRB}. Interestingly, an intensity peak appears in XY polarization at 367 cm$^{-1}$ (see Fig.~\ref{Fig5}) associated with the B$_{g}$(2) mode. This mode is also observed experimentally at 370 cm$^{-1}$ in the XY spectrum, while earlier theoretical calculations based on the Franck-Condon formalism predicted the absence of intensity peaks in the same polarization~\cite{TakuboPRB}. According to their theory, the Mott excitations along the in-plane V-V bond are allowed in the XY configuration, and due to the symmetric nature of the d$_{xy}$ orbital, Mott transitions to the forward and backward direction cancel the intensity~\cite{TakuboPRB}.
%%%%%%%%%%%%%%%%%%%%%%%%%%%%%%%%%%%%%%%%%%%%%%%%%%%%%
\begin{figure}
%\vspace{-1.0cm}
\begin{center}
\includegraphics[width=9.0cm]{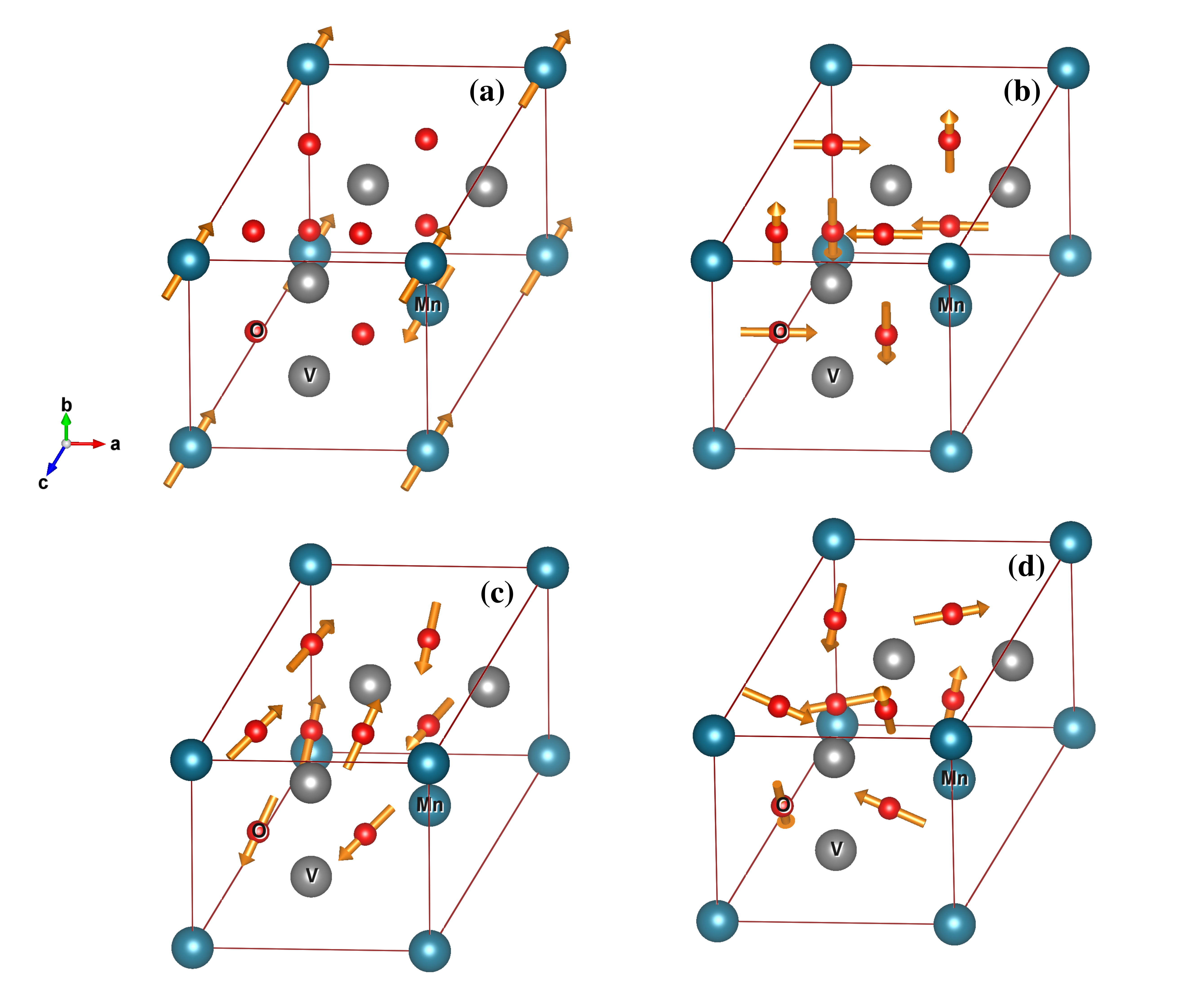}
\caption{\label{Fig6}(Color online) Atomic displacements of B$_{g}$ phonons responsible for the Raman peaks associated with the four B$_{g}$ modes at (a) 181cm$^{-1}$(B$_{g}$(1)), (b) 367cm$^{-1}$(B$_{g}$(2)), (c) 440cm$^{-1}$(B$_{g}$(3)), and (d) 539cm$^{-1}$(B$_{g}$(4)).}
\end{center}
\end{figure}
%%%%%%%%%%%%%%%%%%%%%%%%%%%%%%%%%%%%%%%%%%%%%%%%%%%%%

We discuss the Raman peaks of the specific phonon modes of MnV$_2$O$_4$ in its tetragonal phase, and understand the displacement patterns of the lattice vibrations obtained from {\it ab-initio} phonon calculations. Among the four B$_{g}$ modes, only three B$_g$ modes show the finite intensity of the Raman peaks, while the first one (B$_g$(1)) does not contribute to the Raman scattering process. In Fig.~\ref{Fig6}, we have shown the displacement patterns of these modes. B$_g$(1) mode corresponds to the stretching of the Mn ions along the c-direction (Fig.~\ref{Fig6}(a)), the O ions barely move in this mode. We can see in Fig.~\ref{Fig2}(c),(d) phonons associated with the vibration of Mn ions contribute more to the DOS around that frequency range. The B$_g$(2) (Fig.~\ref{Fig6}(b)) mode, responsible for the Raman peak at 367cm$^{-1}$ comes from the displacements of the O ions along a and b axes while the atomic vibrations of Mn ions are negligibly small. The mode B$_g$(3) and B$_g$(4), responsible for the Raman peaks at 440cm$^{-1}$ and 539cm$^{-1}$ respectively, correspond to the displacements of O atoms as shown in Fig.~\ref{Fig6}(c), and Fig.~\ref{Fig6}(d). We have clearly shown here that the lowest frequency B$_g$ mode emerges due to the phonons associated with Mn ions, and high-frequency B$_g$ modes originate from the atomic vibrations of O ions mainly. However, in all cases, V ions do not move.

\subsection{Spin-phonon coupling}
\begin{figure}
%\vspace{-1.0cm}
\begin{center}
\includegraphics[width=9.0cm]{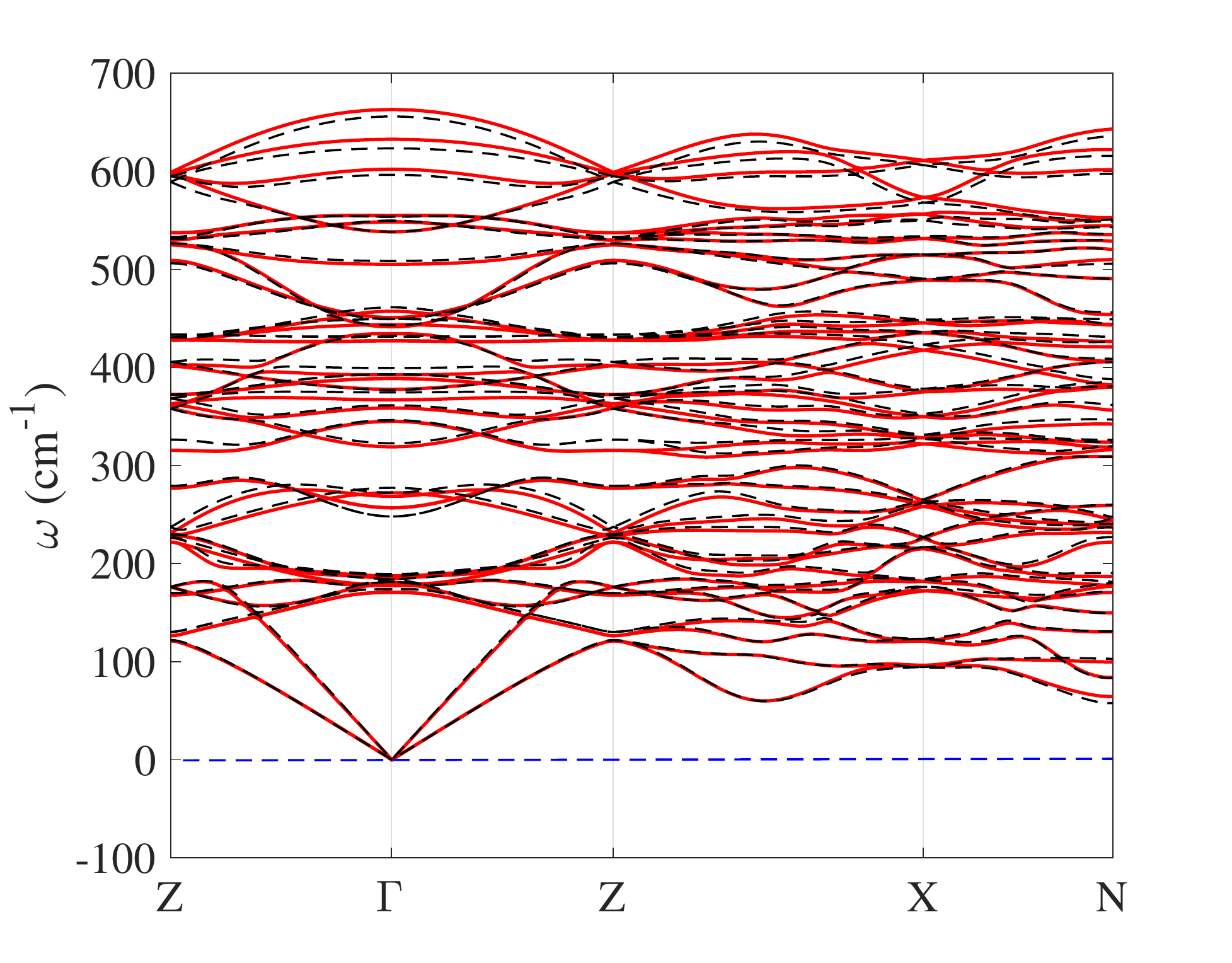}
\caption{\label{Fig7}(Color online) Phonon dispersion of FiM (red solid lines) and FM (black dotted linens) states of MnV$_2$O$_4$ in its tetragonal phase.}
\end{center}
\end{figure}

To understand the interplay between magnetic ordering and phonons, we determine phonons at the zone center and zone boundaries with FiM and FM ordering which gives a measure of spin-phonon coupling. While any spin configuration instead of FM would have been also useful for this purpose, the space group symmetry remains the same in these spin configurations. In Fig.~\ref{Fig7}, we show phonon spectra along high symmetry directions in the Brillouin zone for the FiM and FM spin-ordered states in the tetragonal phase of MnV$_2$O$_4$. In the absence of spin-phonon coupling, hardly any change in phonon spectra is expected in the different magnetic configurations. A strong spin-phonon coupling along the zone center and zone boundary points is  however observed as the phonon frequencies change  with changing magnetic order. In the phonon dispersion (Fig.~\ref{Fig7}), no unstable modes appear at any point in the BZ for either magnetic order.
%%%%%%%%%%%%%%%%%%%%%%%%%%%%%%%%%%%%%%%%%%%%%%
%%%%%%%%%%%%%%%%%%%%%%%%%%%%%%%%%%%%%%%%%%%%%%%%

Heisenberg spin Hamiltonian for MnV$_2$O$_4$ can be written as:
\begin{equation} 
H=\frac{1}{2}\sum\limits_{<ij>}J_{ij}\vec{S_i}\cdot\vec{\sigma_j}-\frac{1}{2}\sum\limits_{<kl>}J_{kl}^{'}\vec{\sigma_k}\cdot\vec{\sigma_l}
\label{eq3}
\end{equation}
$J_{ij}$ is the AFM exchange interaction between $\vec{S_i}$ (Mn spins) and $\vec{\sigma_j}$ (V spins) giving an FiM state, whereas $J_{kl}$ is the FM exchange interaction between V spins. %$\vec{\sigma_k}$ and $\vec{\sigma_l}$. 
Changes in the exchange interactions due to the spin-phonon coupling can be obtained by a Taylor series expansion of $J$ with respect to the amplitude of atomic displacements~\cite{GranadoPRB,KumarPRB},
\begin{equation}
J(\vec{u}_{\nu}^\lambda) = J_0+\vec{u}_{\nu}^\lambda (\nabla_{u_{\nu}}J)+\frac{1}{2}\vec{u}_{\nu}^\lambda (\nabla^2_{u_{\nu\nu'}}J)\vec{u}_{\nu'}^\lambda
\label{eq4}
\end{equation}
%%%%%%%%%%%%%%%%%%%%%%%%%%%%%%%%%%%%
\begin{figure}
%\vspace{-1.0cm}
\begin{center}
\includegraphics[width=7.0cm]{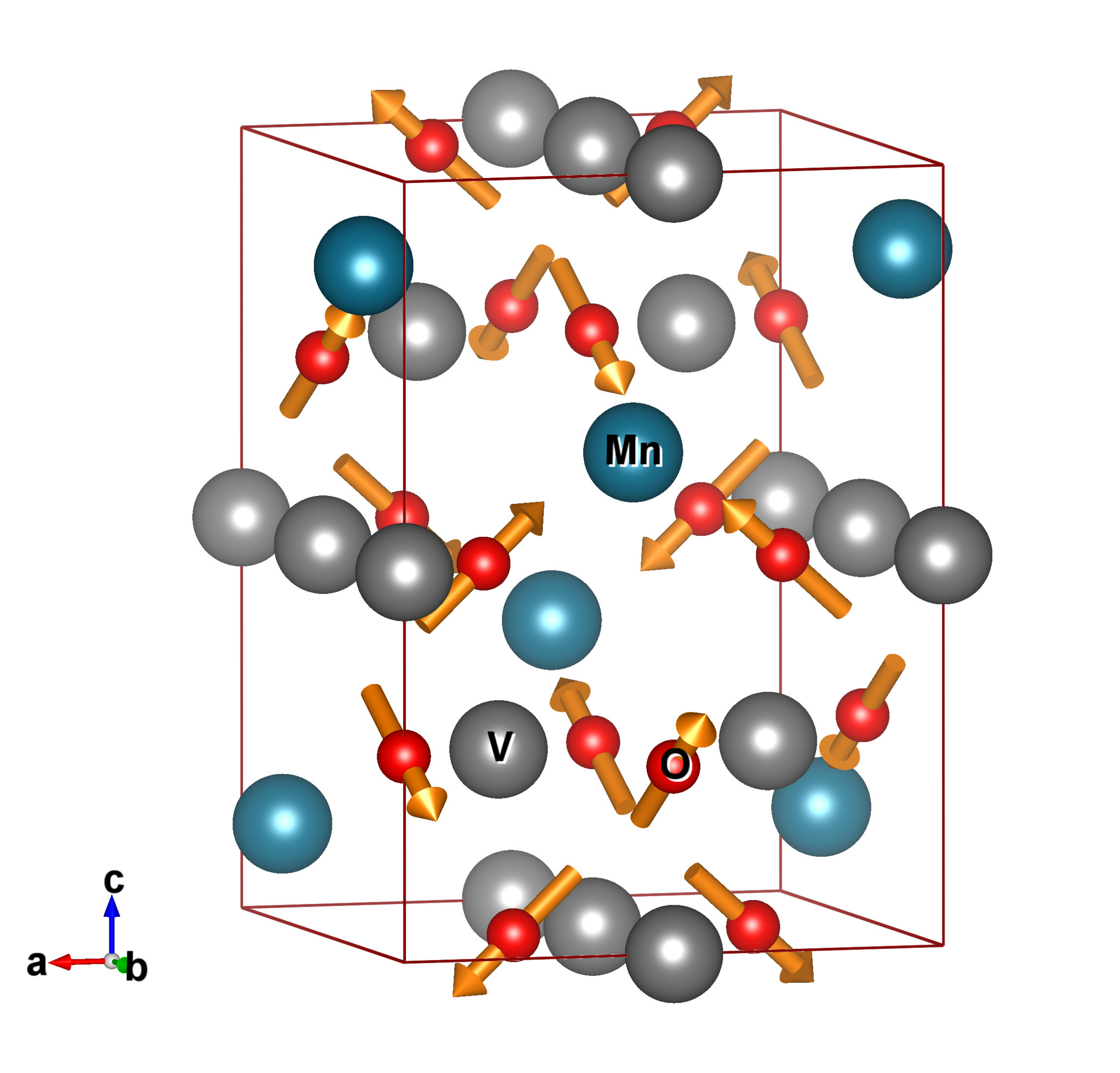}
\caption{\label{Fig8}(Color online) The Hellman-Feynman forces on the Oxygen ions in the FM configuration of the relaxed FiM configuration (crystallographic unit cell). The forces are drawn with golden arrows and give the lowest order spin-phonon coupling.}
\end{center}
\end{figure}
%%%%%%%%%%%%%%%%%%%%%%%%%%%%
Here, $\vec{u}_\nu^\lambda$ is the displacement vector from the equilibrium position of the $\nu^{th}$ ion for the $\lambda$th phonon mode. $J_0$ is the bare spin-spin exchange coupling term, and $\nabla_{u}J$ relates to the forces on atoms arising from the change in magnetic configuration (w.r.t its ground state magnetic order). This term, which is linear in atomic displacements, gives the lowest order coupling between spins and phonons. As seen in Fig.~\ref{Fig8}, these forces are equal and opposite for the pairs of Oxygen atoms, and also have the full symmetry (A$_g$) of the lattice. This could lead to the magneto-elastic anomaly at T$_N$. Shift in the phonon frequency $\Delta_\lambda$ of the $\lambda$th phonon mode due to the change in the magnetic order is related to $\nabla^2_{u}J$~\cite{GranadoPRB} (second-order coupling) by the following expression (the reduced mass and the frequency of the $\lambda$th phonon mode are denoted by $\mu_\lambda$ and $\omega_\lambda$ respectively, and $\hat{u}_{\nu}^\lambda = \vec{u}_{\nu}^\lambda/|\vec{u}_{\nu}^\lambda|$):
\begin{equation}
\Delta_\lambda = \frac{1}{2\mu_\lambda \omega_\lambda}\sum\limits_{\nu}{\hat{u}}_{\nu}^\lambda(\nabla^2_{u_{\nu}}J){\hat{u}}_{\nu}^\lambda = \frac{J''_\lambda}{2\mu_\lambda \omega_\lambda}
\label{eq5}
\end{equation}
This quantity gives an estimate of second-order spin-phonon coupling ($J''_\lambda = \sum\limits_{\nu}{\hat{u}}_{\nu}^\lambda(\nabla^2_{u_{\nu}}J){\hat{u}}_{\nu}^\lambda$~\cite{KumarA}), and hence, large values of $\Delta$ imply strong spin-phonon coupling. Relative change in phonon frequency ($\Delta_{rel}^\lambda$ = $\frac{\omega_{FiM}^\lambda-\omega_{FM}^\lambda}{\omega_{FiM}^\lambda}\times$100\%)~\cite{PaulPRB} due to change in the magnetic order for the $\Gamma$ ({\bf{q}}=$0,0,0$) phonons, as well as for the zone boundary phonons at N ($1/2,0,0$), X ($0,0,1/2$), and Z ($1/2,1/2,-1/2$) points of the BZ are presented in Fig.~\ref{Fig9} (a)-(d).
%%%%%%%%%%%%%%%%%%%%%%%%%%%%%%
\begin{figure}
%\vspace{-1.0cm}
\begin{center}
\includegraphics[width=9.2cm]{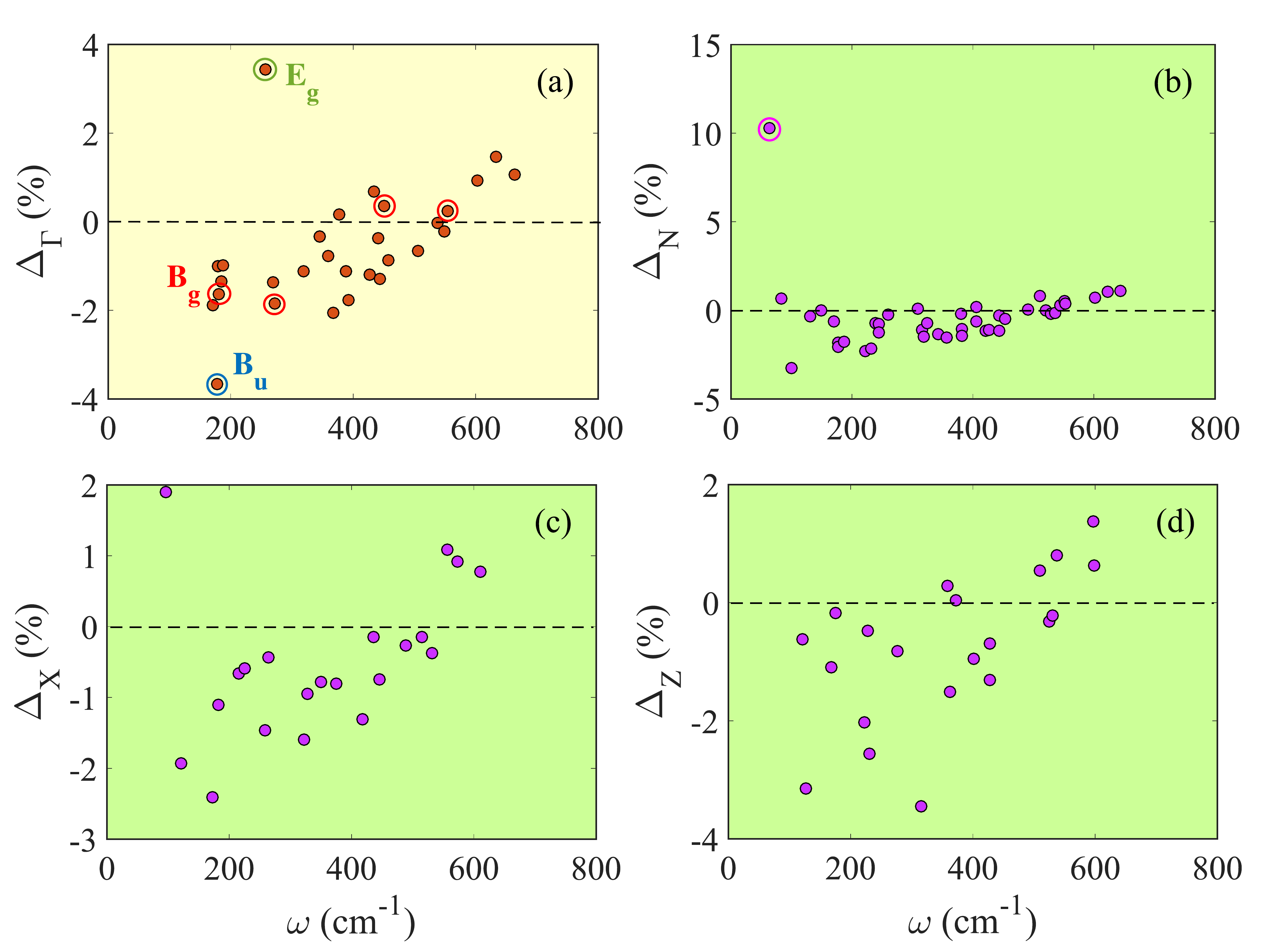}
\caption{\label{Fig9}(Color online) Relative change in phonon frequency ($\Delta_{rel}$ = $\frac{\omega_{FiM}-\omega_{FM}}{\omega_{FiM}}\times$100\%) calculated at the zone center (a) $\Gamma$, and zone boundary points (b) N, (c) X, (d) Z.}
\end{center}
\end{figure}
%%%%%%%%%%%%%%%%%%%%%%%%%%%%%%

In Fig.~\ref{Fig9}(a), two low-frequency (181 cm$^{-1}$, 367 cm$^{-1}$) B$_g$ modes (all B$_g$ modes are marked in red circles) show softening of frequency due to change in spin configuration from FM to FiM. However, two high-frequency B$_g$ modes at 440 cm$^{-1}$ and 549 cm$^{-1}$ show very weak spin-phonon coupling compared to the others. Among all the $\Gamma$ phonons, B$_u$ mode at 178 cm$^{-1}$ (marked by sky-blue circles) shows maximum relative softening ($\sim-4$\%), while 257 cm$^{-1}$ E$_g$ mode (marked by green circles) shows maximum relative hardening due to spin-phonon interaction. Among the zone boundary phonons, 70 cm$^{-1}$ phonon mode (pink circles) (Fig.~\ref{Fig9}(b)) at the N point of the BZ ($1/2,0,0$) shows more than 10\% hardening of frequency due to change in magnetic order from FM to FiM and exhibits strongest spin-phonon coupling across BZ.

\begin{figure}
%\vspace{-1.0cm}
\begin{center}
\includegraphics[width=7.0cm]{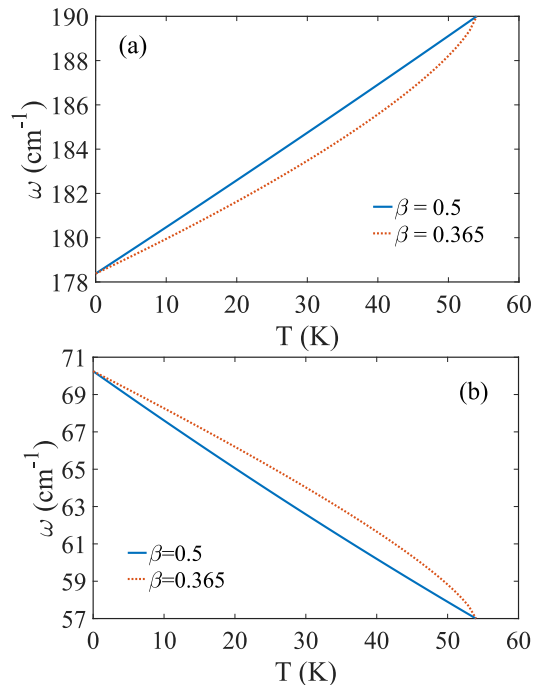}
\caption{\label{Fig10}(Color online) Temperature-dependence of (a) 178 cm$^{-1}$ phonon mode ($\Delta_\lambda=-6$ cm$^{-1}$), and (b) 70 cm$^{-1}$ phonon mode ($\Delta_\lambda=7$ cm$^{-1}$) as obtained from the GL theory. We have considered the critical exponet value~\cite{ZhangEPL} $\beta=0.5$ (mean-field theory) and $\beta=0.365$ (3D-Heisenberg model) respectively in our calculations.}
\end{center}
\end{figure}

To connect our results with experiments~\cite{GleasonPRB} and to understand the observed temperature-dependence of phonon frequencies, we use Ginzburg-Landau (GL) theory. The free-energy of our spin-phonon coupled system can be written as 
\begin{equation}
F = F_0 + a~m^2 + b~m^4 + \sum\limits_{\lambda}(J''_\lambda m^2 v_\lambda^2 + \frac{1}{2} \mu_\lambda {\omega_0}_\lambda^2 v_\lambda^2)  
\label{eq6}
\end{equation}
Here, $a$ and $b$ are usual GL parameters, $m$ is the FiM order parameter, $v_\lambda$ is the amplitude of the $\lambda$th phonon mode, and $\omega_{0\lambda}$ is the high temperature ($T \geq T_N$) phonon frequency. The temperature-dependence of FiM order parameter $m$ is assumed to be $\propto (1-T/T_N)^{\beta}$ for $T<T_N$. It is evident that in the presence of spin-phonon coupling, the phonon frequency will change below $T_N$ as the FiM ordering sets in. The modified phonon frequency is 

\begin{equation}
\omega_\lambda = 2 \Delta_\lambda m^2 \pm \sqrt{{\omega_0}_\lambda^2 + 4\Delta_\lambda^2 m^4} 
\label{eq7}
\end{equation}
Here, the positive solution has been considered, as the other solution gives negative values of $\omega_\lambda$. We have estimated $\Delta_\lambda$ from our DFT calculations and then used it in Eq.~\ref{eq7} to find the temperature-dependence of relevant phonon frequencies (Fig.~\ref{Fig10}) in the tetragonal phase of MnV$_2$O$_4$. We have taken the critical exponent value $\beta=0.5$ (mean-field theory) and $\beta=0.365$ (3D-Heisenberg model) respectively. Zhang {\it et al.}\cite{ZhangEPL} showed that the 3D-Heisenberg model is the best model to describe the critical properties of ferromagnet spinel CdCr$_2$Se$_4$. We have also found that $\beta=0.365$ gives a better temperature-dependence than the mean-field behavior.
%%%%%%%%%%%%%%%%%%%%%%%%%%%%%%
%%%%%%%%%%%%%%%%%%%%%%%%%%%%%%

In Fig.~\ref{Fig10}(a), we see the phonon frequency decreases with decreasing temperature below T$_N$, which is associated with the softening of a phonon mode ($\Delta_\lambda<0$) due to spin-phonon coupling. Interestingly, in the Raman scattering experiments~\cite{TakuboPRB, GleasonPRB}, modes near 178 cm$^{-1}$ show anomalous temperature-dependence as the frequency decreases with decreasing temperature below $T_N$. However, in Fig.~\ref{Fig10}(b) the phonon frequency increases with decreasing temperature below $T_N$ and this is associated with the hardening of a phonon mode ($\Delta_\lambda>0$) due to coupling between spins and phonons. Experimentally, the temperature-dependence of Raman intensity peak near 80 cm$^{-1}$ shows a similar behavior below the FiM transition temperature. Our results agree qualitatively with the experimental observations, and also corroborate the experimental prediction of strong spin-lattice coupling across the BZ in this material.

\section{Conclusions}
\label{conc}
Our first-principles study of phonon dispersions  in two different magnetic orders (FiM and FM) in MnV$_2$O$_4$ show notable spin-phonon coupling. We also find that in the low-temperature tetragonal structure of MnV$_2$O$_4$, correlations are necessary to eliminate unstable modes. Raman intensities for XY and X$'$Y$'$ polarization reveal finite-intensity peaks in the X$'$Y$'$ polarization at two higher frequency B$_g$ modes (B$_g$(3) and B$_g$(4)) and a peak in XY polarization at 367 cm$^{-1}$ which is associated with B$_g$(2) mode. The lowest frequency B$_g$ mode is found to emerge from phonons associated with Mn ions, while high frequency B$_g$ modes originate from O ion vibration. Landau theory analysis reveals the mechanism governing the low-temperature Raman anomalies of MnV$_2$O$_4$. Our results agree qualitatively with recent experiments~\cite{TakuboPRB,GleasonPRB} and reveal that a strong interplay between the lattice and magnetic degrees of freedom is important for understanding the underlying physics of MnV$_2$O$_4$.

\section{Acknowledgments}
D.D. acknowledges Department of Science and Technology (DST), India for an INSPIRE research fellowship. U.V.W. acknowledges support from a J.C. Bose National Fellowship and a Sheikh Saqr Fellowship. A.T. acknowledges research funding from CSIR (India) through the grant number: 03(1373)/16/EMR-II.


\begin{thebibliography}{40}%
\makeatletter
\providecommand \@ifxundefined [1]{%
 \@ifx{#1\undefined}
}%
\providecommand \@ifnum [1]{%
 \ifnum #1\expandafter \@firstoftwo
 \else \expandafter \@secondoftwo
 \fi
}%
\providecommand \@ifx [1]{%
 \ifx #1\expandafter \@firstoftwo
 \else \expandafter \@secondoftwo
 \fi
}%
\providecommand \natexlab [1]{#1}%
\providecommand \enquote  [1]{``#1''}%
\providecommand \bibnamefont  [1]{#1}%
\providecommand \bibfnamefont [1]{#1}%
\providecommand \citenamefont [1]{#1}%
\providecommand \href@noop [0]{\@secondoftwo}%
\providecommand \href [0]{\begingroup \@sanitize@url \@href}%
\providecommand \@href[1]{\@@startlink{#1}\@@href}%
\providecommand \@@href[1]{\endgroup#1\@@endlink}%
\providecommand \@sanitize@url [0]{\catcode `\\12\catcode `\$12\catcode
  `\&12\catcode `\#12\catcode `\^12\catcode `\_12\catcode `\%12\relax}%
\providecommand \@@startlink[1]{}%
\providecommand \@@endlink[0]{}%
\providecommand \url  [0]{\begingroup\@sanitize@url \@url }%
\providecommand \@url [1]{\endgroup\@href {#1}{\urlprefix }}%
\providecommand \urlprefix  [0]{URL }%
\providecommand \Eprint [0]{\href }%
\providecommand \doibase [0]{http://dx.doi.org/}%
\providecommand \selectlanguage [0]{\@gobble}%
\providecommand \bibinfo  [0]{\@secondoftwo}%
\providecommand \bibfield  [0]{\@secondoftwo}%
\providecommand \translation [1]{[#1]}%
\providecommand \BibitemOpen [0]{}%
\providecommand \bibitemStop [0]{}%
\providecommand \bibitemNoStop [0]{.\EOS\space}%
\providecommand \EOS [0]{\spacefactor3000\relax}%
\providecommand \BibitemShut  [1]{\csname bibitem#1\endcsname}%
\let\auto@bib@innerbib\@empty
%</preamble>
\bibitem [{\citenamefont {Tokura}\ and\ \citenamefont
  {Nagaosa}(2000)}]{TokuraSc}%
  \BibitemOpen
  \bibfield  {author} {\bibinfo {author} {\bibfnamefont {Y.}~\bibnamefont
  {Tokura}}\ and\ \bibinfo {author} {\bibfnamefont {N.}~\bibnamefont
  {Nagaosa}},\ }\href {\doibase 10.1126/science.288.5465.462} {\bibfield
  {journal} {\bibinfo  {journal} {Science}\ }\textbf {\bibinfo {volume}
  {288}},\ \bibinfo {pages} {462} (\bibinfo {year} {2000})}\BibitemShut
  {NoStop}%
\bibitem [{\citenamefont {Lee}\ \emph {et~al.}(2010)\citenamefont {Lee},
  \citenamefont {Takagi}, \citenamefont {Louca}, \citenamefont {Matsuda},
  \citenamefont {Ji}, \citenamefont {Ueda}, \citenamefont {Ueda}, \citenamefont
  {Katsufuji}, \citenamefont {Chung}, \citenamefont {Park}, \citenamefont
  {Cheong},\ and\ \citenamefont {Broholm}}]{LeeJPS}%
  \BibitemOpen
  \bibfield  {author} {\bibinfo {author} {\bibfnamefont {S.-H.}\ \bibnamefont
  {Lee}}, \bibinfo {author} {\bibfnamefont {H.}~\bibnamefont {Takagi}},
  \bibinfo {author} {\bibfnamefont {D.}~\bibnamefont {Louca}}, \bibinfo
  {author} {\bibfnamefont {M.}~\bibnamefont {Matsuda}}, \bibinfo {author}
  {\bibfnamefont {S.}~\bibnamefont {Ji}}, \bibinfo {author} {\bibfnamefont
  {H.}~\bibnamefont {Ueda}}, \bibinfo {author} {\bibfnamefont {Y.}~\bibnamefont
  {Ueda}}, \bibinfo {author} {\bibfnamefont {T.}~\bibnamefont {Katsufuji}},
  \bibinfo {author} {\bibfnamefont {J.-H.}\ \bibnamefont {Chung}}, \bibinfo
  {author} {\bibfnamefont {S.}~\bibnamefont {Park}}, \bibinfo {author}
  {\bibfnamefont {S.-W.}\ \bibnamefont {Cheong}}, \ and\ \bibinfo {author}
  {\bibfnamefont {C.}~\bibnamefont {Broholm}},\ }\href {\doibase
  10.1143/JPSJ.79.011004} {\bibfield  {journal} {\bibinfo  {journal} {Journal
  of the Physical Society of Japan}\ }\textbf {\bibinfo {volume} {79}},\
  \bibinfo {pages} {011004} (\bibinfo {year} {2010})}\BibitemShut {NoStop}%
\bibitem [{\citenamefont {Radaelli}(2005)}]{RadaNJP}%
  \BibitemOpen
  \bibfield  {author} {\bibinfo {author} {\bibfnamefont {P.~G.}\ \bibnamefont
  {Radaelli}},\ }\href {\doibase 10.1088/1367-2630/7/1/05} {\bibfield
  {journal} {\bibinfo  {journal} {New Journal of Physics}\ }\textbf {\bibinfo
  {volume} {7}},\ \bibinfo {pages} {53} (\bibinfo {year} {2005})}\BibitemShut
  {NoStop}%
\bibitem [{\citenamefont {Khomskii}\ and\ \citenamefont
  {Mizokawa}(2005)}]{KhomPRL}%
  \BibitemOpen
  \bibfield  {author} {\bibinfo {author} {\bibfnamefont {D.~I.}\ \bibnamefont
  {Khomskii}}\ and\ \bibinfo {author} {\bibfnamefont {T.}~\bibnamefont
  {Mizokawa}},\ }\href {\doibase 10.1103/PhysRevLett.94.156402} {\bibfield
  {journal} {\bibinfo  {journal} {Phys. Rev. Lett.}\ }\textbf {\bibinfo
  {volume} {94}},\ \bibinfo {pages} {156402} (\bibinfo {year}
  {2005})}\BibitemShut {NoStop}%
\bibitem [{\citenamefont {Kato}\ \emph {et~al.}(2012)\citenamefont {Kato},
  \citenamefont {Chern}, \citenamefont {Al-Hassanieh}, \citenamefont
  {Perkins},\ and\ \citenamefont {Batista}}]{KatoPRL}%
  \BibitemOpen
  \bibfield  {author} {\bibinfo {author} {\bibfnamefont {Y.}~\bibnamefont
  {Kato}}, \bibinfo {author} {\bibfnamefont {G.-W.}\ \bibnamefont {Chern}},
  \bibinfo {author} {\bibfnamefont {K.~A.}\ \bibnamefont {Al-Hassanieh}},
  \bibinfo {author} {\bibfnamefont {N.~B.}\ \bibnamefont {Perkins}}, \ and\
  \bibinfo {author} {\bibfnamefont {C.~D.}\ \bibnamefont {Batista}},\ }\href
  {\doibase 10.1103/PhysRevLett.108.247215} {\bibfield  {journal} {\bibinfo
  {journal} {Phys. Rev. Lett.}\ }\textbf {\bibinfo {volume} {108}},\ \bibinfo
  {pages} {247215} (\bibinfo {year} {2012})}\BibitemShut {NoStop}%
\bibitem [{\citenamefont {Tsunetsugu}\ and\ \citenamefont
  {Motome}(2003)}]{TsunePRB}%
  \BibitemOpen
  \bibfield  {author} {\bibinfo {author} {\bibfnamefont {H.}~\bibnamefont
  {Tsunetsugu}}\ and\ \bibinfo {author} {\bibfnamefont {Y.}~\bibnamefont
  {Motome}},\ }\href {\doibase 10.1103/PhysRevB.68.060405} {\bibfield
  {journal} {\bibinfo  {journal} {Phys. Rev. B}\ }\textbf {\bibinfo {volume}
  {68}},\ \bibinfo {pages} {060405} (\bibinfo {year} {2003})}\BibitemShut
  {NoStop}%
\bibitem [{\citenamefont {Choudhury}\ \emph {et~al.}(2014)\citenamefont
  {Choudhury}, \citenamefont {Suzuki}, \citenamefont {Okuyama}, \citenamefont
  {Morikawa}, \citenamefont {Kato}, \citenamefont {Takata}, \citenamefont
  {Kobayashi}, \citenamefont {Kumai}, \citenamefont {Nakao}, \citenamefont
  {Murakami}, \citenamefont {Bremholm}, \citenamefont {Iversen}, \citenamefont
  {Arima}, \citenamefont {Tokura},\ and\ \citenamefont {Taguchi}}]{DebPRB}%
  \BibitemOpen
  \bibfield  {author} {\bibinfo {author} {\bibfnamefont {D.}~\bibnamefont
  {Choudhury}}, \bibinfo {author} {\bibfnamefont {T.}~\bibnamefont {Suzuki}},
  \bibinfo {author} {\bibfnamefont {D.}~\bibnamefont {Okuyama}}, \bibinfo
  {author} {\bibfnamefont {D.}~\bibnamefont {Morikawa}}, \bibinfo {author}
  {\bibfnamefont {K.}~\bibnamefont {Kato}}, \bibinfo {author} {\bibfnamefont
  {M.}~\bibnamefont {Takata}}, \bibinfo {author} {\bibfnamefont
  {K.}~\bibnamefont {Kobayashi}}, \bibinfo {author} {\bibfnamefont
  {R.}~\bibnamefont {Kumai}}, \bibinfo {author} {\bibfnamefont
  {H.}~\bibnamefont {Nakao}}, \bibinfo {author} {\bibfnamefont
  {Y.}~\bibnamefont {Murakami}}, \bibinfo {author} {\bibfnamefont
  {M.}~\bibnamefont {Bremholm}}, \bibinfo {author} {\bibfnamefont {B.~B.}\
  \bibnamefont {Iversen}}, \bibinfo {author} {\bibfnamefont {T.}~\bibnamefont
  {Arima}}, \bibinfo {author} {\bibfnamefont {Y.}~\bibnamefont {Tokura}}, \
  and\ \bibinfo {author} {\bibfnamefont {Y.}~\bibnamefont {Taguchi}},\ }\href
  {\doibase 10.1103/PhysRevB.89.104427} {\bibfield  {journal} {\bibinfo
  {journal} {Phys. Rev. B}\ }\textbf {\bibinfo {volume} {89}},\ \bibinfo
  {pages} {104427} (\bibinfo {year} {2014})}\BibitemShut {NoStop}%
\bibitem [{\citenamefont {Lee}\ \emph {et~al.}(2004)\citenamefont {Lee},
  \citenamefont {Louca}, \citenamefont {Ueda}, \citenamefont {Park},
  \citenamefont {Sato}, \citenamefont {Isobe}, \citenamefont {Ueda},
  \citenamefont {Rosenkranz}, \citenamefont {Zschack}, \citenamefont
  {\'I\~niguez}, \citenamefont {Qiu},\ and\ \citenamefont {Osborn}}]{LeePRL}%
  \BibitemOpen
  \bibfield  {author} {\bibinfo {author} {\bibfnamefont {S.-H.}\ \bibnamefont
  {Lee}}, \bibinfo {author} {\bibfnamefont {D.}~\bibnamefont {Louca}}, \bibinfo
  {author} {\bibfnamefont {H.}~\bibnamefont {Ueda}}, \bibinfo {author}
  {\bibfnamefont {S.}~\bibnamefont {Park}}, \bibinfo {author} {\bibfnamefont
  {T.~J.}\ \bibnamefont {Sato}}, \bibinfo {author} {\bibfnamefont
  {M.}~\bibnamefont {Isobe}}, \bibinfo {author} {\bibfnamefont
  {Y.}~\bibnamefont {Ueda}}, \bibinfo {author} {\bibfnamefont {S.}~\bibnamefont
  {Rosenkranz}}, \bibinfo {author} {\bibfnamefont {P.}~\bibnamefont {Zschack}},
  \bibinfo {author} {\bibfnamefont {J.}~\bibnamefont {\'I\~niguez}}, \bibinfo
  {author} {\bibfnamefont {Y.}~\bibnamefont {Qiu}}, \ and\ \bibinfo {author}
  {\bibfnamefont {R.}~\bibnamefont {Osborn}},\ }\href {\doibase
  10.1103/PhysRevLett.93.156407} {\bibfield  {journal} {\bibinfo  {journal}
  {Phys. Rev. Lett.}\ }\textbf {\bibinfo {volume} {93}},\ \bibinfo {pages}
  {156407} (\bibinfo {year} {2004})}\BibitemShut {NoStop}%
\bibitem [{\citenamefont {MacDougall}\ \emph {et~al.}(2012)\citenamefont
  {MacDougall}, \citenamefont {Garlea}, \citenamefont {Aczel}, \citenamefont
  {Zhou},\ and\ \citenamefont {Nagler}}]{MacPRB}%
  \BibitemOpen
  \bibfield  {author} {\bibinfo {author} {\bibfnamefont {G.~J.}\ \bibnamefont
  {MacDougall}}, \bibinfo {author} {\bibfnamefont {V.~O.}\ \bibnamefont
  {Garlea}}, \bibinfo {author} {\bibfnamefont {A.~A.}\ \bibnamefont {Aczel}},
  \bibinfo {author} {\bibfnamefont {H.~D.}\ \bibnamefont {Zhou}}, \ and\
  \bibinfo {author} {\bibfnamefont {S.~E.}\ \bibnamefont {Nagler}},\ }\href
  {\doibase 10.1103/PhysRevB.86.060414} {\bibfield  {journal} {\bibinfo
  {journal} {Phys. Rev. B}\ }\textbf {\bibinfo {volume} {86}},\ \bibinfo
  {pages} {060414} (\bibinfo {year} {2012})}\BibitemShut {NoStop}%
\bibitem [{\citenamefont {Garlea}\ \emph {et~al.}(2008)\citenamefont {Garlea},
  \citenamefont {Jin}, \citenamefont {Mandrus}, \citenamefont {Roessli},
  \citenamefont {Huang}, \citenamefont {Miller}, \citenamefont {Schultz},\ and\
  \citenamefont {Nagler}}]{GarleaPRL}%
  \BibitemOpen
  \bibfield  {author} {\bibinfo {author} {\bibfnamefont {V.~O.}\ \bibnamefont
  {Garlea}}, \bibinfo {author} {\bibfnamefont {R.}~\bibnamefont {Jin}},
  \bibinfo {author} {\bibfnamefont {D.}~\bibnamefont {Mandrus}}, \bibinfo
  {author} {\bibfnamefont {B.}~\bibnamefont {Roessli}}, \bibinfo {author}
  {\bibfnamefont {Q.}~\bibnamefont {Huang}}, \bibinfo {author} {\bibfnamefont
  {M.}~\bibnamefont {Miller}}, \bibinfo {author} {\bibfnamefont {A.~J.}\
  \bibnamefont {Schultz}}, \ and\ \bibinfo {author} {\bibfnamefont {S.~E.}\
  \bibnamefont {Nagler}},\ }\href {\doibase 10.1103/PhysRevLett.100.066404}
  {\bibfield  {journal} {\bibinfo  {journal} {Phys. Rev. Lett.}\ }\textbf
  {\bibinfo {volume} {100}},\ \bibinfo {pages} {066404} (\bibinfo {year}
  {2008})}\BibitemShut {NoStop}%
\bibitem [{\citenamefont {Nii}\ \emph {et~al.}(2012)\citenamefont {Nii},
  \citenamefont {Sagayama}, \citenamefont {Arima}, \citenamefont {Aoyagi},
  \citenamefont {Sakai}, \citenamefont {Maki}, \citenamefont {Nishibori},
  \citenamefont {Sawa}, \citenamefont {Sugimoto}, \citenamefont {Ohsumi},\ and\
  \citenamefont {Takata}}]{NiiPRB}%
  \BibitemOpen
  \bibfield  {author} {\bibinfo {author} {\bibfnamefont {Y.}~\bibnamefont
  {Nii}}, \bibinfo {author} {\bibfnamefont {H.}~\bibnamefont {Sagayama}},
  \bibinfo {author} {\bibfnamefont {T.}~\bibnamefont {Arima}}, \bibinfo
  {author} {\bibfnamefont {S.}~\bibnamefont {Aoyagi}}, \bibinfo {author}
  {\bibfnamefont {R.}~\bibnamefont {Sakai}}, \bibinfo {author} {\bibfnamefont
  {S.}~\bibnamefont {Maki}}, \bibinfo {author} {\bibfnamefont {E.}~\bibnamefont
  {Nishibori}}, \bibinfo {author} {\bibfnamefont {H.}~\bibnamefont {Sawa}},
  \bibinfo {author} {\bibfnamefont {K.}~\bibnamefont {Sugimoto}}, \bibinfo
  {author} {\bibfnamefont {H.}~\bibnamefont {Ohsumi}}, \ and\ \bibinfo {author}
  {\bibfnamefont {M.}~\bibnamefont {Takata}},\ }\href {\doibase
  10.1103/PhysRevB.86.125142} {\bibfield  {journal} {\bibinfo  {journal} {Phys.
  Rev. B}\ }\textbf {\bibinfo {volume} {86}},\ \bibinfo {pages} {125142}
  (\bibinfo {year} {2012})}\BibitemShut {NoStop}%
\bibitem [{\citenamefont {Tchernyshyov}(2004)}]{TchernyPRL}%
  \BibitemOpen
  \bibfield  {author} {\bibinfo {author} {\bibfnamefont {O.}~\bibnamefont
  {Tchernyshyov}},\ }\href {\doibase 10.1103/PhysRevLett.93.157206} {\bibfield
  {journal} {\bibinfo  {journal} {Phys. Rev. Lett.}\ }\textbf {\bibinfo
  {volume} {93}},\ \bibinfo {pages} {157206} (\bibinfo {year}
  {2004})}\BibitemShut {NoStop}%
\bibitem [{\citenamefont {Di~Matteo}\ \emph {et~al.}(2005)\citenamefont
  {Di~Matteo}, \citenamefont {Jackeli},\ and\ \citenamefont
  {Perkins}}]{MatteoPRB}%
  \BibitemOpen
  \bibfield  {author} {\bibinfo {author} {\bibfnamefont {S.}~\bibnamefont
  {Di~Matteo}}, \bibinfo {author} {\bibfnamefont {G.}~\bibnamefont {Jackeli}},
  \ and\ \bibinfo {author} {\bibfnamefont {N.~B.}\ \bibnamefont {Perkins}},\
  }\href {\doibase 10.1103/PhysRevB.72.020408} {\bibfield  {journal} {\bibinfo
  {journal} {Phys. Rev. B}\ }\textbf {\bibinfo {volume} {72}},\ \bibinfo
  {pages} {020408} (\bibinfo {year} {2005})}\BibitemShut {NoStop}%
\bibitem [{\citenamefont {Suzuki}\ \emph {et~al.}(2007)\citenamefont {Suzuki},
  \citenamefont {Katsumura}, \citenamefont {Taniguchi}, \citenamefont {Arima},\
  and\ \citenamefont {Katsufuji}}]{SuzukiPRL}%
  \BibitemOpen
  \bibfield  {author} {\bibinfo {author} {\bibfnamefont {T.}~\bibnamefont
  {Suzuki}}, \bibinfo {author} {\bibfnamefont {M.}~\bibnamefont {Katsumura}},
  \bibinfo {author} {\bibfnamefont {K.}~\bibnamefont {Taniguchi}}, \bibinfo
  {author} {\bibfnamefont {T.}~\bibnamefont {Arima}}, \ and\ \bibinfo {author}
  {\bibfnamefont {T.}~\bibnamefont {Katsufuji}},\ }\href {\doibase
  10.1103/PhysRevLett.98.127203} {\bibfield  {journal} {\bibinfo  {journal}
  {Phys. Rev. Lett.}\ }\textbf {\bibinfo {volume} {98}},\ \bibinfo {pages}
  {127203} (\bibinfo {year} {2007})}\BibitemShut {NoStop}%
\bibitem [{\citenamefont {Adachi}\ \emph {et~al.}(2005)\citenamefont {Adachi},
  \citenamefont {Suzuki}, \citenamefont {Kato}, \citenamefont {Osaka},
  \citenamefont {Takata},\ and\ \citenamefont {Katsufuji}}]{AdachiPRL}%
  \BibitemOpen
  \bibfield  {author} {\bibinfo {author} {\bibfnamefont {K.}~\bibnamefont
  {Adachi}}, \bibinfo {author} {\bibfnamefont {T.}~\bibnamefont {Suzuki}},
  \bibinfo {author} {\bibfnamefont {K.}~\bibnamefont {Kato}}, \bibinfo {author}
  {\bibfnamefont {K.}~\bibnamefont {Osaka}}, \bibinfo {author} {\bibfnamefont
  {M.}~\bibnamefont {Takata}}, \ and\ \bibinfo {author} {\bibfnamefont
  {T.}~\bibnamefont {Katsufuji}},\ }\href {\doibase
  10.1103/PhysRevLett.95.197202} {\bibfield  {journal} {\bibinfo  {journal}
  {Phys. Rev. Lett.}\ }\textbf {\bibinfo {volume} {95}},\ \bibinfo {pages}
  {197202} (\bibinfo {year} {2005})}\BibitemShut {NoStop}%
\bibitem [{\citenamefont {Sarkar}\ \emph {et~al.}(2009)\citenamefont {Sarkar},
  \citenamefont {Maitra}, \citenamefont {Valent\'{\i}},\ and\ \citenamefont
  {Saha-Dasgupta}}]{SarkarPRL}%
  \BibitemOpen
  \bibfield  {author} {\bibinfo {author} {\bibfnamefont {S.}~\bibnamefont
  {Sarkar}}, \bibinfo {author} {\bibfnamefont {T.}~\bibnamefont {Maitra}},
  \bibinfo {author} {\bibfnamefont {R.}~\bibnamefont {Valent\'{\i}}}, \ and\
  \bibinfo {author} {\bibfnamefont {T.}~\bibnamefont {Saha-Dasgupta}},\ }\href
  {\doibase 10.1103/PhysRevLett.102.216405} {\bibfield  {journal} {\bibinfo
  {journal} {Phys. Rev. Lett.}\ }\textbf {\bibinfo {volume} {102}},\ \bibinfo
  {pages} {216405} (\bibinfo {year} {2009})}\BibitemShut {NoStop}%
\bibitem [{\citenamefont {Dey}\ \emph {et~al.}(2016)\citenamefont {Dey},
  \citenamefont {Maitra},\ and\ \citenamefont {Taraphder}}]{DeyPRB}%
  \BibitemOpen
  \bibfield  {author} {\bibinfo {author} {\bibfnamefont {D.}~\bibnamefont
  {Dey}}, \bibinfo {author} {\bibfnamefont {T.}~\bibnamefont {Maitra}}, \ and\
  \bibinfo {author} {\bibfnamefont {A.}~\bibnamefont {Taraphder}},\ }\href
  {\doibase 10.1103/PhysRevB.93.195133} {\bibfield  {journal} {\bibinfo
  {journal} {Phys. Rev. B}\ }\textbf {\bibinfo {volume} {93}},\ \bibinfo
  {pages} {195133} (\bibinfo {year} {2016})}\BibitemShut {NoStop}%
\bibitem [{\citenamefont {Takubo}\ \emph {et~al.}(2011)\citenamefont {Takubo},
  \citenamefont {Kubota}, \citenamefont {Suzuki}, \citenamefont {Kanzaki},
  \citenamefont {Miyahara}, \citenamefont {Furukawa},\ and\ \citenamefont
  {Katsufuji}}]{TakuboPRB}%
  \BibitemOpen
  \bibfield  {author} {\bibinfo {author} {\bibfnamefont {K.}~\bibnamefont
  {Takubo}}, \bibinfo {author} {\bibfnamefont {R.}~\bibnamefont {Kubota}},
  \bibinfo {author} {\bibfnamefont {T.}~\bibnamefont {Suzuki}}, \bibinfo
  {author} {\bibfnamefont {T.}~\bibnamefont {Kanzaki}}, \bibinfo {author}
  {\bibfnamefont {S.}~\bibnamefont {Miyahara}}, \bibinfo {author}
  {\bibfnamefont {N.}~\bibnamefont {Furukawa}}, \ and\ \bibinfo {author}
  {\bibfnamefont {T.}~\bibnamefont {Katsufuji}},\ }\href {\doibase
  10.1103/PhysRevB.84.094406} {\bibfield  {journal} {\bibinfo  {journal} {Phys.
  Rev. B}\ }\textbf {\bibinfo {volume} {84}},\ \bibinfo {pages} {094406}
  (\bibinfo {year} {2011})}\BibitemShut {NoStop}%
\bibitem [{\citenamefont {Gleason}\ \emph {et~al.}(2014)\citenamefont
  {Gleason}, \citenamefont {Byrum}, \citenamefont {Gim}, \citenamefont
  {Thaler}, \citenamefont {Abbamonte}, \citenamefont {MacDougall},
  \citenamefont {Martin}, \citenamefont {Zhou},\ and\ \citenamefont
  {Cooper}}]{GleasonPRB}%
  \BibitemOpen
  \bibfield  {author} {\bibinfo {author} {\bibfnamefont {S.~L.}\ \bibnamefont
  {Gleason}}, \bibinfo {author} {\bibfnamefont {T.}~\bibnamefont {Byrum}},
  \bibinfo {author} {\bibfnamefont {Y.}~\bibnamefont {Gim}}, \bibinfo {author}
  {\bibfnamefont {A.}~\bibnamefont {Thaler}}, \bibinfo {author} {\bibfnamefont
  {P.}~\bibnamefont {Abbamonte}}, \bibinfo {author} {\bibfnamefont {G.~J.}\
  \bibnamefont {MacDougall}}, \bibinfo {author} {\bibfnamefont {L.~W.}\
  \bibnamefont {Martin}}, \bibinfo {author} {\bibfnamefont {H.~D.}\
  \bibnamefont {Zhou}}, \ and\ \bibinfo {author} {\bibfnamefont {S.~L.}\
  \bibnamefont {Cooper}},\ }\href {\doibase 10.1103/PhysRevB.89.134402}
  {\bibfield  {journal} {\bibinfo  {journal} {Phys. Rev. B}\ }\textbf {\bibinfo
  {volume} {89}},\ \bibinfo {pages} {134402} (\bibinfo {year}
  {2014})}\BibitemShut {NoStop}%
\bibitem [{\citenamefont {Kohn}\ and\ \citenamefont {Sham}(1965)}]{KohnShamPR}%
  \BibitemOpen
  \bibfield  {author} {\bibinfo {author} {\bibfnamefont {W.}~\bibnamefont
  {Kohn}}\ and\ \bibinfo {author} {\bibfnamefont {L.~J.}\ \bibnamefont
  {Sham}},\ }\href {\doibase 10.1103/PhysRev.140.A1133} {\bibfield  {journal}
  {\bibinfo  {journal} {Phys. Rev.}\ }\textbf {\bibinfo {volume} {140}},\
  \bibinfo {pages} {A1133} (\bibinfo {year} {1965})}\BibitemShut {NoStop}%
\bibitem [{\citenamefont {Hohenberg}\ and\ \citenamefont
  {Kohn}(1964)}]{HohenKohnPR}%
  \BibitemOpen
  \bibfield  {author} {\bibinfo {author} {\bibfnamefont {P.}~\bibnamefont
  {Hohenberg}}\ and\ \bibinfo {author} {\bibfnamefont {W.}~\bibnamefont
  {Kohn}},\ }\href {\doibase 10.1103/PhysRev.136.B864} {\bibfield  {journal}
  {\bibinfo  {journal} {Phys. Rev.}\ }\textbf {\bibinfo {volume} {136}},\
  \bibinfo {pages} {B864} (\bibinfo {year} {1964})}\BibitemShut {NoStop}%
\bibitem [{\citenamefont {Kresse}\ and\ \citenamefont
  {Furthm\"uller}(1996)}]{KresseVASP1}%
  \BibitemOpen
  \bibfield  {author} {\bibinfo {author} {\bibfnamefont {G.}~\bibnamefont
  {Kresse}}\ and\ \bibinfo {author} {\bibfnamefont {J.}~\bibnamefont
  {Furthm\"uller}},\ }\href {\doibase 10.1103/PhysRevB.54.11169} {\bibfield
  {journal} {\bibinfo  {journal} {Phys. Rev. B}\ }\textbf {\bibinfo {volume}
  {54}},\ \bibinfo {pages} {11169} (\bibinfo {year} {1996})}\BibitemShut
  {NoStop}%
\bibitem [{\citenamefont {Kresse}\ and\ \citenamefont
  {Furthmüller}(1996)}]{KresseVASP2}%
  \BibitemOpen
  \bibfield  {author} {\bibinfo {author} {\bibfnamefont {G.}~\bibnamefont
  {Kresse}}\ and\ \bibinfo {author} {\bibfnamefont {J.}~\bibnamefont
  {Furthmüller}},\ }\href {\doibase
  https://doi.org/10.1016/0927-0256(96)00008-0} {\bibfield  {journal} {\bibinfo
   {journal} {Computational Materials Science}\ }\textbf {\bibinfo {volume}
  {6}},\ \bibinfo {pages} {15 } (\bibinfo {year} {1996})}\BibitemShut {NoStop}%
\bibitem [{\citenamefont {Bl\"ochl}(1994)}]{BlochPAW}%
  \BibitemOpen
  \bibfield  {author} {\bibinfo {author} {\bibfnamefont {P.~E.}\ \bibnamefont
  {Bl\"ochl}},\ }\href {\doibase 10.1103/PhysRevB.50.17953} {\bibfield
  {journal} {\bibinfo  {journal} {Phys. Rev. B}\ }\textbf {\bibinfo {volume}
  {50}},\ \bibinfo {pages} {17953} (\bibinfo {year} {1994})}\BibitemShut
  {NoStop}%
\bibitem [{\citenamefont {Kresse}\ and\ \citenamefont
  {Joubert}(1999)}]{KressePAW}%
  \BibitemOpen
  \bibfield  {author} {\bibinfo {author} {\bibfnamefont {G.}~\bibnamefont
  {Kresse}}\ and\ \bibinfo {author} {\bibfnamefont {D.}~\bibnamefont
  {Joubert}},\ }\href {\doibase 10.1103/PhysRevB.59.1758} {\bibfield  {journal}
  {\bibinfo  {journal} {Phys. Rev. B}\ }\textbf {\bibinfo {volume} {59}},\
  \bibinfo {pages} {1758} (\bibinfo {year} {1999})}\BibitemShut {NoStop}%
\bibitem [{\citenamefont {Perdew}\ \emph {et~al.}(1996)\citenamefont {Perdew},
  \citenamefont {Burke},\ and\ \citenamefont {Ernzerhof}}]{GGAPBE}%
  \BibitemOpen
  \bibfield  {author} {\bibinfo {author} {\bibfnamefont {J.~P.}\ \bibnamefont
  {Perdew}}, \bibinfo {author} {\bibfnamefont {K.}~\bibnamefont {Burke}}, \
  and\ \bibinfo {author} {\bibfnamefont {M.}~\bibnamefont {Ernzerhof}},\ }\href
  {\doibase 10.1103/PhysRevLett.77.3865} {\bibfield  {journal} {\bibinfo
  {journal} {Phys. Rev. Lett.}\ }\textbf {\bibinfo {volume} {77}},\ \bibinfo
  {pages} {3865} (\bibinfo {year} {1996})}\BibitemShut {NoStop}%
\bibitem [{\citenamefont {Dudarev}\ \emph {et~al.}(1998)\citenamefont
  {Dudarev}, \citenamefont {Botton}, \citenamefont {Savrasov}, \citenamefont
  {Humphreys},\ and\ \citenamefont {Sutton}}]{DudarevPRB}%
  \BibitemOpen
  \bibfield  {author} {\bibinfo {author} {\bibfnamefont {S.~L.}\ \bibnamefont
  {Dudarev}}, \bibinfo {author} {\bibfnamefont {G.~A.}\ \bibnamefont {Botton}},
  \bibinfo {author} {\bibfnamefont {S.~Y.}\ \bibnamefont {Savrasov}}, \bibinfo
  {author} {\bibfnamefont {C.~J.}\ \bibnamefont {Humphreys}}, \ and\ \bibinfo
  {author} {\bibfnamefont {A.~P.}\ \bibnamefont {Sutton}},\ }\href {\doibase
  10.1103/PhysRevB.57.1505} {\bibfield  {journal} {\bibinfo  {journal} {Phys.
  Rev. B}\ }\textbf {\bibinfo {volume} {57}},\ \bibinfo {pages} {1505}
  (\bibinfo {year} {1998})}\BibitemShut {NoStop}%
\bibitem [{\citenamefont {Hubbard}(1963)}]{Hubbard}%
  \BibitemOpen
  \bibfield  {author} {\bibinfo {author} {\bibfnamefont {J.}~\bibnamefont
  {Hubbard}},\ }\href {\doibase 10.1098/rspa.1963.0204} {\bibfield  {journal}
  {\bibinfo  {journal} {Proceedings of the Royal Society of London A:
  Mathematical, Physical and Engineering Sciences}\ }\textbf {\bibinfo {volume}
  {276}},\ \bibinfo {pages} {238} (\bibinfo {year} {1963})}\BibitemShut
  {NoStop}%
\bibitem [{\citenamefont {Togo}\ and\ \citenamefont {Tanaka}(2015)}]{Phonopy}%
  \BibitemOpen
  \bibfield  {author} {\bibinfo {author} {\bibfnamefont {A.}~\bibnamefont
  {Togo}}\ and\ \bibinfo {author} {\bibfnamefont {I.}~\bibnamefont {Tanaka}},\
  }\href {\doibase https://doi.org/10.1016/j.scriptamat.2015.07.021} {\bibfield
   {journal} {\bibinfo  {journal} {Scripta Materialia}\ }\textbf {\bibinfo
  {volume} {108}},\ \bibinfo {pages} {1 } (\bibinfo {year} {2015})}\BibitemShut
  {NoStop}%
\bibitem [{\citenamefont {Parlinski}(2013)}]{PHONON}%
  \BibitemOpen
  \bibfield  {author} {\bibinfo {author} {\bibfnamefont {K.}~\bibnamefont
  {Parlinski}},\ }\href@noop {} {}\bibinfo {organization} {PHONON Software},\
  \bibinfo {address} {Krak\'ow} (\bibinfo {year} {2013})\BibitemShut {NoStop}%
\bibitem [{\citenamefont {Ray}\ and\ \citenamefont {Waghmare}(2008)}]{RayPRB}%
  \BibitemOpen
  \bibfield  {author} {\bibinfo {author} {\bibfnamefont {N.}~\bibnamefont
  {Ray}}\ and\ \bibinfo {author} {\bibfnamefont {U.~V.}\ \bibnamefont
  {Waghmare}},\ }\href {\doibase 10.1103/PhysRevB.77.134112} {\bibfield
  {journal} {\bibinfo  {journal} {Phys. Rev. B}\ }\textbf {\bibinfo {volume}
  {77}},\ \bibinfo {pages} {134112} (\bibinfo {year} {2008})}\BibitemShut
  {NoStop}%
\bibitem [{\citenamefont {Cardona}\ and\ \citenamefont
  {G\"untherodt}(1982)}]{Cardona}%
  \BibitemOpen
  \bibfield  {author} {\bibinfo {author} {\bibfnamefont {M.}~\bibnamefont
  {Cardona}}\ and\ \bibinfo {author} {\bibfnamefont {G.}~\bibnamefont
  {G\"untherodt}},\ }\href@noop {} {\emph {\bibinfo {title} {Light Scattering
  in Solids II}}}\ (\bibinfo  {publisher} {Springer-Verlag},\ \bibinfo
  {address} {Berlin},\ \bibinfo {year} {1982})\BibitemShut {NoStop}%
\bibitem [{\citenamefont {Br\"uesch}(1986)}]{Bru}%
  \BibitemOpen
  \bibfield  {author} {\bibinfo {author} {\bibfnamefont {P.}~\bibnamefont
  {Br\"uesch}},\ }\href@noop {} {\emph {\bibinfo {title} {Phonons: Theory and
  Experiments II}}}\ (\bibinfo  {publisher} {Springer-Verlag},\ \bibinfo
  {address} {Berlin},\ \bibinfo {year} {1986})\BibitemShut {NoStop}%
\bibitem [{\citenamefont {Umari}\ \emph {et~al.}(2001)\citenamefont {Umari},
  \citenamefont {Pasquarello},\ and\ \citenamefont {Dal~Corso}}]{UmariPRB}%
  \BibitemOpen
  \bibfield  {author} {\bibinfo {author} {\bibfnamefont {P.}~\bibnamefont
  {Umari}}, \bibinfo {author} {\bibfnamefont {A.}~\bibnamefont {Pasquarello}},
  \ and\ \bibinfo {author} {\bibfnamefont {A.}~\bibnamefont {Dal~Corso}},\
  }\href {\doibase 10.1103/PhysRevB.63.094305} {\bibfield  {journal} {\bibinfo
  {journal} {Phys. Rev. B}\ }\textbf {\bibinfo {volume} {63}},\ \bibinfo
  {pages} {094305} (\bibinfo {year} {2001})}\BibitemShut {NoStop}%
\bibitem [{\citenamefont {Ceriotti}\ \emph {et~al.}(2006)\citenamefont
  {Ceriotti}, \citenamefont {Pietrucci},\ and\ \citenamefont
  {Bernasconi}}]{CeriottiPRB}%
  \BibitemOpen
  \bibfield  {author} {\bibinfo {author} {\bibfnamefont {M.}~\bibnamefont
  {Ceriotti}}, \bibinfo {author} {\bibfnamefont {F.}~\bibnamefont {Pietrucci}},
  \ and\ \bibinfo {author} {\bibfnamefont {M.}~\bibnamefont {Bernasconi}},\
  }\href {\doibase 10.1103/PhysRevB.73.104304} {\bibfield  {journal} {\bibinfo
  {journal} {Phys. Rev. B}\ }\textbf {\bibinfo {volume} {73}},\ \bibinfo
  {pages} {104304} (\bibinfo {year} {2006})}\BibitemShut {NoStop}%
\bibitem [{\citenamefont {Granado}\ \emph {et~al.}(1999)\citenamefont
  {Granado}, \citenamefont {Garc\'{\i}a}, \citenamefont {Sanjurjo},
  \citenamefont {Rettori}, \citenamefont {Torriani}, \citenamefont {Prado},
  \citenamefont {S\'anchez}, \citenamefont {Caneiro},\ and\ \citenamefont
  {Oseroff}}]{GranadoPRB}%
  \BibitemOpen
  \bibfield  {author} {\bibinfo {author} {\bibfnamefont {E.}~\bibnamefont
  {Granado}}, \bibinfo {author} {\bibfnamefont {A.}~\bibnamefont
  {Garc\'{\i}a}}, \bibinfo {author} {\bibfnamefont {J.~A.}\ \bibnamefont
  {Sanjurjo}}, \bibinfo {author} {\bibfnamefont {C.}~\bibnamefont {Rettori}},
  \bibinfo {author} {\bibfnamefont {I.}~\bibnamefont {Torriani}}, \bibinfo
  {author} {\bibfnamefont {F.}~\bibnamefont {Prado}}, \bibinfo {author}
  {\bibfnamefont {R.~D.}\ \bibnamefont {S\'anchez}}, \bibinfo {author}
  {\bibfnamefont {A.}~\bibnamefont {Caneiro}}, \ and\ \bibinfo {author}
  {\bibfnamefont {S.~B.}\ \bibnamefont {Oseroff}},\ }\href {\doibase
  10.1103/PhysRevB.60.11879} {\bibfield  {journal} {\bibinfo  {journal} {Phys.
  Rev. B}\ }\textbf {\bibinfo {volume} {60}},\ \bibinfo {pages} {11879}
  (\bibinfo {year} {1999})}\BibitemShut {NoStop}%
\bibitem [{\citenamefont {Kumar}\ \emph
  {et~al.}(2012{\natexlab{a}})\citenamefont {Kumar}, \citenamefont {Bera},
  \citenamefont {Muthu}, \citenamefont {Shirodkar}, \citenamefont {Saha},
  \citenamefont {Shireen}, \citenamefont {Sundaresan}, \citenamefont
  {Waghmare}, \citenamefont {Sood},\ and\ \citenamefont {Rao}}]{KumarPRB}%
  \BibitemOpen
  \bibfield  {author} {\bibinfo {author} {\bibfnamefont {P.}~\bibnamefont
  {Kumar}}, \bibinfo {author} {\bibfnamefont {A.}~\bibnamefont {Bera}},
  \bibinfo {author} {\bibfnamefont {D.~V.~S.}\ \bibnamefont {Muthu}}, \bibinfo
  {author} {\bibfnamefont {S.~N.}\ \bibnamefont {Shirodkar}}, \bibinfo {author}
  {\bibfnamefont {R.}~\bibnamefont {Saha}}, \bibinfo {author} {\bibfnamefont
  {A.}~\bibnamefont {Shireen}}, \bibinfo {author} {\bibfnamefont
  {A.}~\bibnamefont {Sundaresan}}, \bibinfo {author} {\bibfnamefont {U.~V.}\
  \bibnamefont {Waghmare}}, \bibinfo {author} {\bibfnamefont {A.~K.}\
  \bibnamefont {Sood}}, \ and\ \bibinfo {author} {\bibfnamefont {C.~N.~R.}\
  \bibnamefont {Rao}},\ }\href {\doibase 10.1103/PhysRevB.85.134449} {\bibfield
   {journal} {\bibinfo  {journal} {Phys. Rev. B}\ }\textbf {\bibinfo {volume}
  {85}},\ \bibinfo {pages} {134449} (\bibinfo {year}
  {2012}{\natexlab{a}})}\BibitemShut {NoStop}%
\bibitem [{\citenamefont {Kumar}\ \emph
  {et~al.}(2012{\natexlab{b}})\citenamefont {Kumar}, \citenamefont {Fennie},\
  and\ \citenamefont {Rabe}}]{KumarA}%
  \BibitemOpen
  \bibfield  {author} {\bibinfo {author} {\bibfnamefont {A.}~\bibnamefont
  {Kumar}}, \bibinfo {author} {\bibfnamefont {C.~J.}\ \bibnamefont {Fennie}}, \
  and\ \bibinfo {author} {\bibfnamefont {K.~M.}\ \bibnamefont {Rabe}},\ }\href
  {\doibase 10.1103/PhysRevB.86.184429} {\bibfield  {journal} {\bibinfo
  {journal} {Phys. Rev. B}\ }\textbf {\bibinfo {volume} {86}},\ \bibinfo
  {pages} {184429} (\bibinfo {year} {2012}{\natexlab{b}})}\BibitemShut
  {NoStop}%
\bibitem [{\citenamefont {Paul}\ \emph {et~al.}(2017)\citenamefont {Paul},
  \citenamefont {Chatterjee}, \citenamefont {Roy}, \citenamefont {Midya},
  \citenamefont {Mandal}, \citenamefont {Grover},\ and\ \citenamefont
  {Tyagi}}]{PaulPRB}%
  \BibitemOpen
  \bibfield  {author} {\bibinfo {author} {\bibfnamefont {B.}~\bibnamefont
  {Paul}}, \bibinfo {author} {\bibfnamefont {S.}~\bibnamefont {Chatterjee}},
  \bibinfo {author} {\bibfnamefont {A.}~\bibnamefont {Roy}}, \bibinfo {author}
  {\bibfnamefont {A.}~\bibnamefont {Midya}}, \bibinfo {author} {\bibfnamefont
  {P.}~\bibnamefont {Mandal}}, \bibinfo {author} {\bibfnamefont
  {V.}~\bibnamefont {Grover}}, \ and\ \bibinfo {author} {\bibfnamefont {A.~K.}\
  \bibnamefont {Tyagi}},\ }\href {\doibase 10.1103/PhysRevB.95.054103}
  {\bibfield  {journal} {\bibinfo  {journal} {Phys. Rev. B}\ }\textbf {\bibinfo
  {volume} {95}},\ \bibinfo {pages} {054103} (\bibinfo {year}
  {2017})}\BibitemShut {NoStop}%
\bibitem [{\citenamefont {Zhang}\ \emph {et~al.}(2010)\citenamefont {Zhang},
  \citenamefont {Fan}, \citenamefont {Li}, \citenamefont {Li}, \citenamefont
  {Ling}, \citenamefont {Qu}, \citenamefont {Tong}, \citenamefont {Tan},\ and\
  \citenamefont {Zhang}}]{ZhangEPL}%
  \BibitemOpen
  \bibfield  {author} {\bibinfo {author} {\bibfnamefont {L.}~\bibnamefont
  {Zhang}}, \bibinfo {author} {\bibfnamefont {J.}~\bibnamefont {Fan}}, \bibinfo
  {author} {\bibfnamefont {L.}~\bibnamefont {Li}}, \bibinfo {author}
  {\bibfnamefont {R.}~\bibnamefont {Li}}, \bibinfo {author} {\bibfnamefont
  {L.}~\bibnamefont {Ling}}, \bibinfo {author} {\bibfnamefont {Z.}~\bibnamefont
  {Qu}}, \bibinfo {author} {\bibfnamefont {W.}~\bibnamefont {Tong}}, \bibinfo
  {author} {\bibfnamefont {S.}~\bibnamefont {Tan}}, \ and\ \bibinfo {author}
  {\bibfnamefont {Y.}~\bibnamefont {Zhang}},\ }\href {\doibase
  10.1209/0295-5075/91/57001} {\bibfield  {journal} {\bibinfo  {journal} {EPL
  (Europhysics Letters)}\ }\textbf {\bibinfo {volume} {91}},\ \bibinfo {pages}
  {57001} (\bibinfo {year} {2010})}\BibitemShut {NoStop}%
\end{thebibliography}
\end{document}